%
%
%

\input harvmac

\input epsf

\noblackbox
\newcount\figno
\figno=0
\def\fig#1#2#3#4{
\par\begingroup\parindent=0pt\leftskip=1cm\rightskip=1cm\parindent=0pt
\baselineskip=11pt
\global\advance\figno by 1
\midinsert
\epsfxsize=#3
\epsfysize=#4
\centerline{\epsfbox{#2}}
\vskip 12pt
{\bf Figure\ \the\figno: } #1\par
\endinsert\endgroup\par
}
\def\doublefig#1#2#3#4#5{
\par\begingroup\parindent=0pt\leftskip=1cm\rightskip=1cm\parindent=0pt
\baselineskip=11pt
\global\advance\figno by 1
\midinsert
\centerline{\hbox{\epsfxsize=#4
\epsfysize=#5
\epsfbox{#2} \hskip 1cm
\epsfxsize=#4
\epsfysize=#5
\epsfbox{#3}}}
\vskip 12pt
{\bf Figure\ \the\figno: } #1\par
\endinsert\endgroup\par
}
\def\figlabel#1{\xdef#1{\the\figno}}
\def\tilde{\widetilde}
 \font\teneusm=eusm10 \font\seveneusm=eusm7 \font\fiveeusm=eusm5
\newfam\eusmfam
\textfont\eusmfam=\teneusm \scriptfont\eusmfam=\seveneusm
\scriptscriptfont\eusmfam=\fiveeusm
\def\eusm#1{{\fam\eusmfam\relax#1}}


\lref\BekensteinJP{
  J.~D.~Bekenstein,
  ``A Universal Upper Bound On The Entropy To Energy Ratio For Bounded
  Systems,''
  Phys.\ Rev.\  D {\bf 23}, 287 (1981).
}

\lref\Rankin{R. Rankin, ``The Zeros of Certain Poincar\'e
Series,'' Compositio Mathmatica, {\bf 46} (1982) 255-272. }

\lref\leeyangone{
  C.~N.~Yang and T.~D.~Lee,
  ``Statistical Theory of Equations of State and Phase Transitions. I: Theory
  of Condensation,''
  Phys.\ Rev.\  {\bf 87}, 404 (1952).
}

\lref\CarlipNV{
  S.~Carlip,
  Class.\ Quant.\ Grav.\  {\bf 17}, 4175 (2000)
  [arXiv:gr-qc/0005017].
}

\lref\leeyangtwo{
 T.~D.~Lee and C.~N.~Yang,
  ``Statistical Theory of Equations of State and Phase Transitions. II: Lattice
  Gas and Ising Model,''
  Phys.\ Rev.\  {\bf 87}, 410 (1952).
}

\lref\kanekonote{
M.~Kaneko, private communication.}

\lref\jdt{
  S.~Deser, R.~Jackiw and G.~'t Hooft,
  ``Three-Dimensional Einstein Gravity: Dynamics Of Flat Space,''
  Annals Phys.\  {\bf 152}, 220 (1984).
}

\lref\WittenKT{
  E.~Witten,
  ``Three-Dimensional Gravity Revisited,''
  arXiv:0706.3359 [hep-th].
}

\lref\farey{R. Dijkgraaf, J. Maldacena, G. Moore, and E. Verlinde,
``A Black Hole Farey Tail,'' hep-th/0005003.}

 \lref\malda{J. Maldacena, ``The Large $N$ Limit Of
Superconformal Field Theories And Supergravity,'' Adv. Theor.
Math. Phys. {\bf 2} (1998) 231-252, hep-th/9711200.}

\lref\AKN{ T. Asai, M. Kaneko, H. Ninomiya, ``Zeros of Certain
Modular Functions and an Application,'' Commentarii Math. Univ.
Sancti Pauli, vol. 46-1 (1997)  93-101.}

\lref\Apostol{T. Apostol, {\it Modular Functions and Dirichlet Series
in Number Theory}, Springer Verlag, 1990.}

\lref\ManschotZB{
  J.~Manschot,
  ``${\rm AdS}_3$ Partition Functions Reconstructed,''
  arXiv:0707.1159 [hep-th].
}

\lref\coadj{E. Witten, ``Coadjoint Orbits Of The Virasoro Group,''
Commun. Math. Phys. {\bf 114} (1988) 1-53.}

\lref\carliptwo{  S.~Carlip and C.~Teitelboim,
  ``Aspects Of Black Hole Quantum Mechanics And Thermodynamics In
  (2+1)-Dimensions,''
  Phys.\ Rev.\  D {\bf 51}, 622 (1995)
  [arXiv:gr-qc/9405070].
}

\lref\italian{  A.~A.~Bytsenko, L.~Vanzo and S.~Zerbini,
  ``Quantum Correction to the Entropy of the (2+1)-Dimensional Black Hole,''
  Phys.\ Rev.\  D {\bf 57}, 4917 (1998)
  [arXiv:gr-qc/9710106].
}

\lref\brownhen{J. D. Brown and M. Henneaux, ``Central Charges In
The Canonical Realization Of Asymptotical Symmetries: An Example
{}From Three-Dimensional Gravity,'' Commun. Math. Phys. {\bf 104}
(1986) 207-226.}

\lref\ban{M. Ba\~{n}ados, C. Teitelboim, and J. Zanelli, ``The
Black Hole In Three-Dimensional Spacetime,'' Phys. Rev. Lett. {\bf
69} (1992) 1849-1851, hep-th/9204099.}

\lref\bantwo{ M.~Ba\~{n}ados, M.~Henneaux, C.~Teitelboim and J.~Zanelli,
  ``Geometry of the (2+1) Black Hole,''
  Phys.\ Rev.\  D {\bf 48}, 1506 (1993)
  [arXiv:gr-qc/9302012].
}

\lref\rabadan{  M.~Kleban, M.~Porrati and R.~Rabadan,
  ``Poincare Recurrences and Topological Diversity,''
  JHEP {\bf 0410}, 030 (2004)
  [arXiv:hep-th/0407192].
}

 \lref\strom{A. Strominger, ``Black Hole Entropy
{}From Near Horizon Microstates,'' JHEP 9802:009 (1998),
hep-th/9712251.}

\lref\GaiottoXH{
  D.~Gaiotto and X.~Yin,
  ``Genus Two Partition Functions of Extremal Conformal Field Theories,''
  JHEP {\bf 0708}, 029 (2007)
  [arXiv:0707.3437 [hep-th]].
}

\lref\MaldacenaBW{
  J.~M.~Maldacena and A.~Strominger,
  ``AdS(3) Black Holes and a Stringy Exclusion Principle,''
  JHEP {\bf 9812}, 005 (1998)
  [arXiv:hep-th/9804085].
}

\lref\bounded{ G.~W.~Gibbons, S.~W.~Hawking and M.~J.~Perry,
  ``Path Integrals And The Indefiniteness Of The Gravitational Action,''
  Nucl.\ Phys.\  B {\bf 138}, 141 (1978).
}

\lref\hawkingpage{ S.~W.~Hawking and D.~N.~Page,
  ``Thermodynamics Of Black Holes In Anti-De Sitter Space,''
  Commun.\ Math.\ Phys.\  {\bf 87}, 577 (1983).
}

\lref\GaberdielVE{
  M.~R.~Gaberdiel,
  ``Constraints on Extremal Self-dual CFTs,''
  arXiv:0707.4073 [hep-th].
}
\lref\YinGV{
  X.~Yin,
  ``Partition Functions of Three-Dimensional Pure Gravity,''
  arXiv:0710.2129 [hep-th].
}

\lref\townsend{A. Ach\'{u}carro and P. Townsend,  ``A Chern-Simons
Action for Three-Dimensional
anti-De Sitter Supergravity Theories,'' Phys. Lett. {\bf B180} (1986) 89.}%
\lref\witten{E. Witten, ``(2+1)-Dimensional Gravity as an Exactly
Soluble System,'' Nucl. Phys. {\bf B311} (1988) 46.}%

 \lref\carlip{S. Carlip, ``Conformal Field Theory,
$(2+1)$-Dimensional Gravity, and the BTZ Black Hole,''
gr-qc/0503022.}

\lref\BrownNW{
  J.~D.~Brown and M.~Henneaux,
  ``Central Charges in the Canonical Realization of Asymptotic Symmetries: An
  Example from Three-Dimensional Gravity,''
  Commun.\ Math.\ Phys.\  {\bf 104}, 207 (1986).
}

\lref\jackiw{
 S.~Deser and R.~Jackiw,
  ``Three-Dimensional Cosmological Gravity: Dynamics Of Constant Curvature,''
  Annals Phys.\  {\bf 153}, 405 (1984).
}

\lref\raysinger{ D.~B.~Ray and I.~M.~Singer,
  ``Analytic Torsion For Complex Manifolds,''
  Annals Math.\  {\bf 98}, 154 (1973).
}

\lref\Iwaniec{
H. Iwaniec, {\it Spectral Methods of Automorphic Forms},
American Mathematical Society, 2002.
}

\Title{\vbox{\baselineskip12pt\hbox{arXiv:yymm.nnnn} }}{Quantum
Gravity Partition Functions In Three Dimensions}\vskip .1in

\centerline{ Alexander Maloney$^{1}$\footnote{}{email:
maloney@physics.mcgill.ca, witten@ias.edu} and Edward
Witten$^{2}$} \vskip .1in

\

\centerline{
${}^{1}$McGill Physics Department, 3600 rue University, Montr\'eal, QC H3A 2T8, Canada}

\centerline{
${}^{2}$School of Natural Sciences, Institute for Advanced Study, Einstein Dr., Princeton, NJ 08540}

\vskip .5in \centerline{\bf Abstract} { We consider pure
three-dimensional quantum gravity with a negative cosmological
constant. The sum of known contributions to the partition function
from classical geometries can be computed exactly, including
quantum corrections. However, the result is not physically
sensible, and if the model does exist, there are some additional
contributions.
One possibility is that the theory may
have long strings and a continuous spectrum.
Another possibility is that complex geometries need to
be included, possibly leading to a holomorphically factorized
partition function.
We analyze the subleading corrections to the Bekenstein-Hawking entropy and show that these can be correctly reproduced in such a holomorphically factorized theory.  We also consider the Hawking-Page phase transition
between a thermal gas and a black hole and show that it is a phase
transition of Lee-Yang type, associated with a condensation of
zeros in the complex temperature plane.  Finally, we analyze pure
three-dimensional supergravity, with similar results.} \vskip .3in

\smallskip
\Date{}

\listtoc
\writetoc
\

\newsec{Introduction}

This paper is devoted to describing some explicit computations
relevant to three-dimensional pure quantum gravity with negative
cosmological constant.  The classical action can be written
\eqn\harrigo{I={1\over 16\pi G}\int d^3x\sqrt g\left(R+{2\over
\ell^2}\right).} Three-dimensional gravity with a cosmological
constant was first studied in \jackiw.  Some early milestones in
the study of this theory, both of them special to the case of
negative cosmological constant, have been the construction of an
asymptotic Virasoro symmetry \brownhen\ and the recognition that
the theory admits black holes \refs{\ban,\bantwo}.   The
asymptotic Virasoro symmetry is  now understood as part of the
structure of a dual two-dimensional conformal field theory \malda\
and this structure together with modular invariance can be used to
determine the entropy of a black hole of asymptotically large mass
\strom. For a recent review of the extensive work on this subject,
with references, see \carlip.

\def\AdS{{\rm AdS}}
None of the results mentioned in the last paragraph are special to
the case of {\it pure} three-dimensional gravity, without
additional fields.  Our intent here, however, is to study this
minimal theory, with the goal of computing its exact energy levels
in a spacetime that is asymptotic at spatial infinity to the
classical ``vacuum'' of three-dimensional Anti de Sitter space,
$\AdS_3$.  As always in General Relativity, a well-posed problem
is obtained by specifying what the world should look like at
spatial infinity -- in this case we ask that it should coincide
with $\AdS_3$ at infinity -- and then analyzing all possible
``interiors.''  Our problem in this paper  is to compute the
precise quantum energy levels that arise.  We perform a
computation based on known concepts, and in a sense that we will
explain, this computation is not successful.  The reasons for this failure are not
clear and we consider several hypotheses.

 It is convenient to summarize the spectrum of energy levels
in the form of a trace $\Tr\,\exp(-\beta H)$, where $H$ is the
Hamiltonian and $\beta$ is a positive real number (or more
generally a complex number with positive real part). As usual in
General Relativity, the Hamiltonian is defined via the ADM
procedure in terms of the leading behavior at spatial infinity of
the correction to the pure $\AdS_3$ metric. It was indeed by
carefully examining this procedure that the asymptotic Virasoro
symmetry (which generalizes the obvious conservation laws such as
conservation of energy) was discovered \brownhen.

There is also a conserved angular momentum $J$ which generates a
rotation at infinity of the asymptotically $\AdS_3$ spacetime and
commutes with $H$.  Consequently, one can introduce an additional
parameter $\theta$ and try to compute a more general partition
function: \eqn\moreg{Z(\beta,\theta)=\Tr\,\exp(-\beta H-i\theta
J).}

This partition function is naturally computed via a Euclidean path
integral.  According to the standard recipe, the integral is taken
over Euclidean three-geometries that are conformal at infinity to
a two-torus $\Sigma$ with modular parameter
$\tau=\theta/2\pi+i\beta$. We write $|dz|^2$ for a flat metric on
this torus, where $z$ is a complex coordinate subject to the
identifications $z\to z+1$, $z\to z+\tau$, and introduce another
coordinate $u>0$, such that conformal infinity will be the region
$u\to 0$.  Then the metric should behave for $u\to 0$  like
\eqn\looklike{ds^2={|dz|^2+du^2\over u^2}} plus  subleading terms.

The Euclidean path integral also has a natural interpretation in
the dual two-dimensional CFT \malda: it is the genus one partition
function of this dual theory, on the two-dimensional surface
$\Sigma$.  In fact, by definition, the Hamiltonian and momentum
$H$ and $P$ of the dual CFT coincide with $H$ and $J$, the
Hamiltonian and angular momentum of the original theory in
$\AdS_3$.  This dual interpretation is informative, but is not
really needed to motivate the computation that we perform.

This Euclidean path integral is a formal recipe, for various
reasons.  One problem is that, in general, the Euclidean quantum
gravity path integral is not convergent because the action is
not bounded below \bounded.  The only known way to deal with this
problem is to expand around a classical solution; in doing so, one
can obtain a meaningful result at least in perturbation theory.
There is no clearly established claim in the literature that
topologies that do not admit classical solutions do not contribute
to the Euclidean path integral.  But there is also no known method
to evaluate their contributions.

It turns out that in the present case, one can describe the
classical solutions completely.  This is explained in Sec. 2.1.
The classical solutions are precisely the ones considered in \MaldacenaBW\
and in \farey\ (in studying the elliptic genus of certain $\AdS_3$ models
derived from string theory).  Moreover, perturbation theory around
a classical solution terminates with the one-loop term, and that
term can be easily evaluated by adapting the arguments of
\brownhen.  This is the content of Sec. 2.2.

We are therefore in a position to write down the complete sum of
{\it known} contributions to the path integral.  This is done in
Sec. 3.  The sum turns out to require some regularization, and we
use what we believe is a natural regularization, analogous to zeta
function regularization \raysinger.  We ultimately obtain an
explicit, though complicated, formula for the sum of known
contributions to $Z(\beta,\theta)$.  However -- and this is our
main result -- the sum is not physically sensible: it cannot be
written as $\Tr\,\exp(-\beta H-i\theta J)$ for any commuting
operators $H,J$ in a Hilbert space. (According to \YinGV, this possibility was also conjectured by S. Minwalla.)

We do not know the correct interpretation of this result. In Sec.
4, we discuss some possibilities.  One possibility is that the
{\it minimal} theory that we postulate here actually does not
exist.   There are many quantum theories that do exist that look
semiclassically like three-dimensional gravity coupled to
additional fields; indeed, there are a plethora of known
string-derived models, such as the ones studied in \farey.
However, it is not clear\foot{The problem may be analogous to
trying to define a minimal string theory in four dimensions. There
are many theories that macroscopically are four-dimensional string
theories, such as  theories obtained by compactification to four
dimensions from the critical dimension, or gauge theories with
flux tubes or vortex lines, but it is plausible that there is
nothing that should sensibly be called a minimal four-dimensional
quantum string theory.} that there exists nonperturbatively a
minimal theory that one would want to call pure gravity.  One
possible interpretation of our result is to indicate that this
minimal theory does not exist.

The other possibility, broadly speaking, is that there are
additional contributions to the path integral beyond the known
ones that we evaluate in Secs.  2 and 3.  To be concrete, we
consider in Sec. 4 two scenarios.  One is that the minimal theory
of three-dimensional gravity, in addition to the known BTZ black
holes and Brown-Henneaux boundary excitations, also describes cosmic
strings. The motivation for this proposal is that known models of
three-dimensional  quantum gravity, such as the string-based
models considered in \farey, do always have cosmic strings.
Perhaps this is also true for the ``minimal'' theory, if it
exists.

The second scenario is that in addition to the real saddle points
that are classified and evaluated in Secs. 2 and 3, an exact
description of the theory should also include complex saddle
points.  We describe a specific scenario in which the inclusion of
complex saddle points leads to the holomorphically factorized
partition function that was proposed (based on highly speculative
reasoning) in \WittenKT.  This involves a doubled sum over saddle points
similar to what is
assumed  in \refs{\YinGV, \ManschotZB}.  The resulting partition function is consistent
with an interpretation as $\Tr\,\exp(-\beta H-i\theta J)$ with
some Hilbert space operators $H,J$.

Though we are not able to put this proposal in a convincing form,
we do uncover one interesting fact, which concerns the
semiclassical behavior of the partition function (assumed to be
holomorphically factorized and extremal) as the gravitational
coupling $G$ goes to zero with fixed AdS radius $\ell$. In this
semiclassical limit, the partition function $Z=\Tr\,\exp(-\beta
H-i\theta J)$ is always dominated, as long as $\beta$ and $\theta$
are real, by a real saddle point corresponding to a real geometry.
The partition function can be dominated by a classical saddle
point, but only if one asks a more exotic question with complex
values of $\beta$ and $\theta$.

In Secs. 5 and 6 we discuss questions of black hole physics in three dimensions.  In section 5 we discuss black hole entropy: a theory of pure quantum gravity, assuming it exists, should give a proper microscopic accounting of the Bekenstein-Hawking entropy of the BTZ black hole.  In the semi-classical limit this entropy is just the horizon area.  The computations of section 2, however, allow us to determine the perturbative corrections to this semi-classical result, which are described in section 5.1.  In section 5.2 we show that an infinite series of corrections can be reproduced in a holomorphically factorized theory.

In investigating these holomorphically factorized theories, we noticed an additional interesting
phenomenon that is the subject of Sec. 6.  Three-dimensional
quantum gravity, as a function of $\beta$, exhibits in the
semi-classical limit the Hawking-Page phase transition
\hawkingpage\ between a thermal gas (in this case, a gas of
Brown-Henneaux boundary excitations) and a black hole.  At first
sight, it is puzzling how such a phase transition can be
compatible with holomorphic factorization.  We show, however, that
this question has a natural answer, in terms of a condensation on
the phase boundary of Lee-Yang zeroes of the partition function.

Finally, in Sec. 7, we extend our analysis to pure ${\cal N}=1$
supergravity, obtaining results similar to what we find in the
bosonic case.

\newsec{Known Contributions To The Path Integral}

Our task in this section is two-fold.  The first step is to
classify Euclidean solutions of Einstein's equations, with
negative cosmological constant, with the asymptotic behavior described
in the introduction.  We write $M$ for the three-dimensional
spacetime, and $\Sigma$ for its conformal boundary; as in the
introduction, $\Sigma$ is a Riemann surface of genus 1.  We assume
that $M$ is smooth, that the metric on $M$ is complete, and that
$\Sigma$ is the only ``end'' of $M$. According to the standard
logic of Euclidean quantum gravity, known contributions to the
path integral have these properties; it may be appropriate
physically to relax them, but we do not know how.

We proceed in two steps.  In Sec. 2.1, we describe the possible
choices of $M$.  The result is actually well-known in the theory
of hyperbolic three-manifolds.  In Sec. 2.2, we evaluate the
contribution to the partition function of a particular $M$. The
sum over different choices of $M$ is postponed to Sec. 3.  In Sec.
2.3, we describe the general form of the partition function in a
theory with finite entropy.

\subsec{Classification Of Solutions}

\input amssym.tex
\def\C{\Bbb{C}}
\def\Z{\Bbb{Z}}
\def\CP{\Bbb{CP}}
\def\R{\Bbb{R}}

The automorphism group of $\AdS_3$ is $SO(3,1)$, which is the same
as $SL(2,\C)/\Z_2$. We may write the metric on a dense open subset
of $\AdS_{3}$ as\foot{We set $\ell=1$ unless otherwise indicated.}
\eqn\adsmet{ ds^{2} = {|dz|^{2} + du^{2}\over
u^{2}},~~~~~u>0,~~~~~z\in\C .} If we combine the $(z,u)$
coordinates into a single quaternion $y=z+ju$, the action of an
element $\left( a~b\atop c~d\right) \in SL(2,\C)$ can be written
succinctly as \eqn\quatform{ y \to \left(ay + b\right) \left( cy +
d\right)^{-1} .}In this expression the element $\left(-1~0\atop0~-1\right)\in SL(2,\C)$ acts trivially, so \quatform\ actually describes the action of $SL(2,\C)/\Z_{2}$ on $\AdS_{3}$. In general, any classical solution $M$ of
three-dimensional gravity with negative cosmological constant
takes the form $\AdS_3'/\Gamma$, where $\Gamma$ is a discrete
subgroup of $SO(3,1)$ and $\AdS_3'$ is the part of $\AdS_3$ on
which $\Gamma$ acts discretely.

The conformal boundary of $M$ can be constructed as follows. First
of all, the conformal boundary of $\AdS_3$ is a two-sphere, which
one can think of as $\CP^1$, acted on by $SL(2,\C)$ in the usual
way.  This $\CP^{1}$ may be regarded as the complex $z$-plane in \adsmet\
at $u\to0$ plus a point
at infinity.  From \quatform, one can see that $SL(2,\C)$ acts on this $\CP^{1}$ in the
familiar fashion
\eqn\familiar{z\to {az+b\over cz+d}.}

To construct the conformal boundary of $M=\AdS_3'/\Gamma$, one
first throws away a certain subset of $\CP^1$ in a neighborhood of
which the discrete group $\Gamma$ acts badly. This set is closed,
so its complement is an open subset $U\subset \CP^1$.

The discrete group $\Gamma$ acts freely on $U$, and the conformal boundary of $M$ is
the quotient \eqn\quotilent{\Sigma=U/\Gamma.} Since the action of
$SL(2,\C)$ preserves the holomorphic structure of $\CP^1$,
$U/\Gamma$ carries a natural holomorphic structure and the
isomorphism \quotilent\ is valid holomorphically, not just
topologically.

 Given this, let us
investigate the condition for $\Sigma$ to be of genus 1. $\Sigma$
is topologically a two-torus, so its fundamental group is
$\pi_1(\Sigma)=\Z\oplus \Z$.  It follows from eqn. \quotilent\
that the fundamental group of $U$ is a subgroup of the fundamental
group of $\Sigma$.  Possible subgroups  $\pi_1(U)\subset\Z\times
\Z$ are of three types:

{\it (i)} $\pi_1(U)$ may be a subgroup of $\pi_1(\Sigma)$ of
finite index, isomorphic to $\Z\oplus \Z$.  (A special case is
$\pi_1(U)=\pi_1(\Sigma)$.)

{\it (ii)} $\pi_1(U)$ may consist of multiples of a single
non-zero vector $x\in \pi_1(\Sigma)$, in which case $\pi_1(U)\cong
\Z$.

{\it (iii)} $\pi_1(U)$ may be trivial.

In case {\it (i)}, $U$ is a finite cover of $\Sigma$, and
therefore is itself a Riemann surface of genus 1.  However, a
Riemann surface of genus 1 is not isomorphic to an open subset of
$\CP^1$.  So case {\it (i)} cannot arise.

\bigskip\noindent{\it Cusp Geometry}

In case {\it (iii)}, $U$ is the universal cover of $\Sigma$, and
so is isomorphic to $\C$ (or  $\R^2$).  The holomorphic structure
of $\C$ is unique up to isomorphism.  $\C$ is isomorphic
holomorphically to an open subset of $\CP^{1}$ in essentially only
one way: it is the complement of one point, say the point at
$z=\infty$. The subgroup of $SL(2,\C)$ that leaves fixed the point
at infinity consists of the triangular matrices
\eqn\triang{\left(\matrix{\lambda & w \cr 0 &
\lambda^{-1}}\right).} The point $z\in\C$ corresponds to
$\left(\matrix{z\cr 1\cr}\right)\in\CP^1$, and a triangular matrix
acts by $z\to \lambda^2z+\lambda w$.

Since $U$ is simply-connected, $\Gamma$ must be isomorphic to
$\Z\oplus \Z$ (in order to get the right fundamental group for
$\Sigma$).  Any discrete group of triangular matrices that is
isomorphic to $\Z\oplus\Z$ is generated by two strictly triangular
matrices \eqn\riang{\left(\matrix{1 & a \cr 0 &
1}\right),~~\left(\matrix{1 & b \cr 0 & 1}\right),} where $a$ and
$b$ are complex numbers that are linearly independent over $\R$.
(If $a$ and $b$ are linearly dependent over $\R$, then the group
generated by these matrices is not discrete, since a suitable
linear combination $ma+nb$, for $m,n\in \Z$, can be arbitrarily
small.) Up to conjugacy by a diagonal matrix, the only invariant
of such a group is the ratio $b/a$. Therefore, we can reduce to
the case $a=1$, $b=\tau$, and moreover by taking $b\to -b$ (which
does not affect the group generated by the two matrices) we can
assume that ${\rm Im}\,\tau>0$.

We have therefore arrived precisely at the group of symmetries
\eqn\itro{z\to z+m+n\tau,~m,n\in \Z} of the complex $z$-plane. The
quotient is a genus 1 surface $\Sigma$ with an arbitrary
$\tau$-parameter. The three-manifold $M=\AdS_3/\Gamma$ can be
described very concretely and is a standard example. The $(u,z)$
coordinates in \adsmet\ cover precisely the subspace $\AdS_3'$ on
which $\Gamma$ acts ``nicely.'' So $M=\AdS_3'/\Gamma$ is given by
the metric \adsmet, now subject to the identification \itro.

But $M$ does not obey the conditions described at the beginning of
this section.  $M$ is smooth and has a complete Einstein metric.
However, in addition to the ``end'' at $u=0$, which is the one
required by our boundary condition, $M$ also has a second ``end''
at $u=\infty$. Our problem was to classify Einstein manifolds with
only one end; $M$ does not qualify.

The second end of $M$ is one at which the metric of $\Sigma$
collapses to zero (rather than blowing up, as it does for $u\to
0$).  An end of this kind is known as a ``cusp.''  In the theory
of hyperbolic three-manifolds, one often considers hyperbolic
metrics with such cusps. One of the main results about them is
that, under suitable conditions, such a metric can be slightly
perturbed so as to eliminate the cusp.  In the present context,
this gives the sort of metric that we will describe momentarily.

Physically, we cannot be certain that omitting the case of the
``cusp'' is the correct thing to do.  However, restricting to
cuspless metrics appears to be the right thing to do in known
examples of $\AdS_3$ theories.  And pragmatically, we believe that
from what is known of three-dimensional gravity, it would be
difficult to give a sensible procedure for evaluating the
contribution of the spacetime with the cusp.  The reason for the
last statement is that non-trivial one-cycles (loops in $\Sigma$)
become sub-Planckian in length near a cusp, so a semiclassical
treatment is not valid.

\bigskip\noindent{\it Semiclassical Geometries}

The last case to consider  -- case {\it (ii)} -- is that the fundamental group of $U$ is
$\Z$.  This means that topologically $U$ is $\R\times S^1$ ($S^1$
is a circle).  The holomorphic structure of $U$ is then uniquely
determined to be that of the $z$-plane minus a point, which we may
as well take to be the point at $z=0$.  The subgroup of triangular
matrices that preserve the point $z=0$ is simply the group of
diagonal matrices.

Therefore $\Gamma$ is a discrete subgroup of the group of diagonal
matrices.  It cannot be a finite group (or the first Betti
number of $\Sigma=U/\Gamma$ would be 1, while the desired value is
2).  This being so, there are essentially two cases to consider.

First, $\Gamma$ may be isomorphic to $\Z$, generated by a matrix of the
form \eqn\genby{W=\left(\matrix{q & 0 \cr 0 & q^{-1}}\right) \in SL(2,\C).}
By exchanging the two eigenvalues we can assume that
$|q|<1$. (If $|q|=1$, then either $q$ is a root of unity and the
group generated by $W$ is a finite group, or $q$ is not a root of
unity and the subgroup of $SL(2,\C)$ generated by $W$ is not
discrete.) Alternatively, $\Gamma$ may be isomorphic to $\Z\times \Z_n$,
generated by $W$ together with \eqn\enby{Y=\left(\matrix{\exp(2\pi
i/n) & 0\cr 0&\exp(-2\pi i/n)}\right),} with some integer $n$.

Let us first consider the case that $\Gamma=\Z$.  Then
$\Sigma=U/\Gamma$ is obtained from the complex $z$-plane by
throwing away the point $z=0$ and dividing by the group generated
by $W$.  It is convenient to write $z=\exp(2\pi iw)$, so that $w$
is defined modulo \eqn\defby{w\to w+1} and $W$ acts by
\eqn\nefby{w\to w+{\log q\over 2\pi i}.}

The quotient of the $w$-plane by \defby\ and \nefby\ is a Riemann
surface of genus 1, as required.
The complex modulus of this surface is $\tau = {\log q \over 2\pi i}$, i.e.
it is given by $q=e^{2\pi i \tau}$.
More generally, however, the modulus of this Riemann surface
is defined only up to $\tau\to (a\tau+b)/(c\tau+d)$ with
integers $a,b,c,d$ obeying $ad-bc=1.$  Therefore, we will get an
equivalent Riemann surface if \eqn\zefby{q=\exp(2\pi
i(a\tau+b)/(c\tau+d))} for such $a,b,c,d$.

One might conclude, therefore, that we get a three-manifold
obeying the required conditions for every choice of $a,b,c,d$.
This is not quite the case, for two reasons.  First, an overall
sign change of $a,b,c,d$ does not affect $q$ or the associated
three-manifold.  Second,  once $c$ and $d$ are given, $a$ and $b$
are uniquely determined by $ad-bc=1$ up to shifts of the form
$(a,b)\to (a,b)+t(c,d)$, $t\in \Z$.  Under this transformation,
$q$ as defined in eqn. \zefby\ is invariant. So the possible
three-manifolds really only depend on the choice of the pair $c,d$
of relatively prime integers, up to sign.  For each such pair, we
find integers $a,b$ such that $ad-bc=1$, and identify $q$ via eqn.
\zefby. This gives a manifold that we will call $M_{c,d}$. This
family of manifolds were first discussed in the context of
three-dimensional gravity in \MaldacenaBW.

\bigskip\noindent{\it The Geometry of $M_{c,d}$}

\fig{ a) An infinite cylinder representing $\AdS_3$.  The boundary
of the cylinder represents conformal infinity; time translations
act by vertical shifts.  b) A slice of height determined by
$\beta$. The manifold $M_{0,1}$ is built by gluing together the
top and bottom, after a rotation that identifies the boundary
points marked by solid dots.} {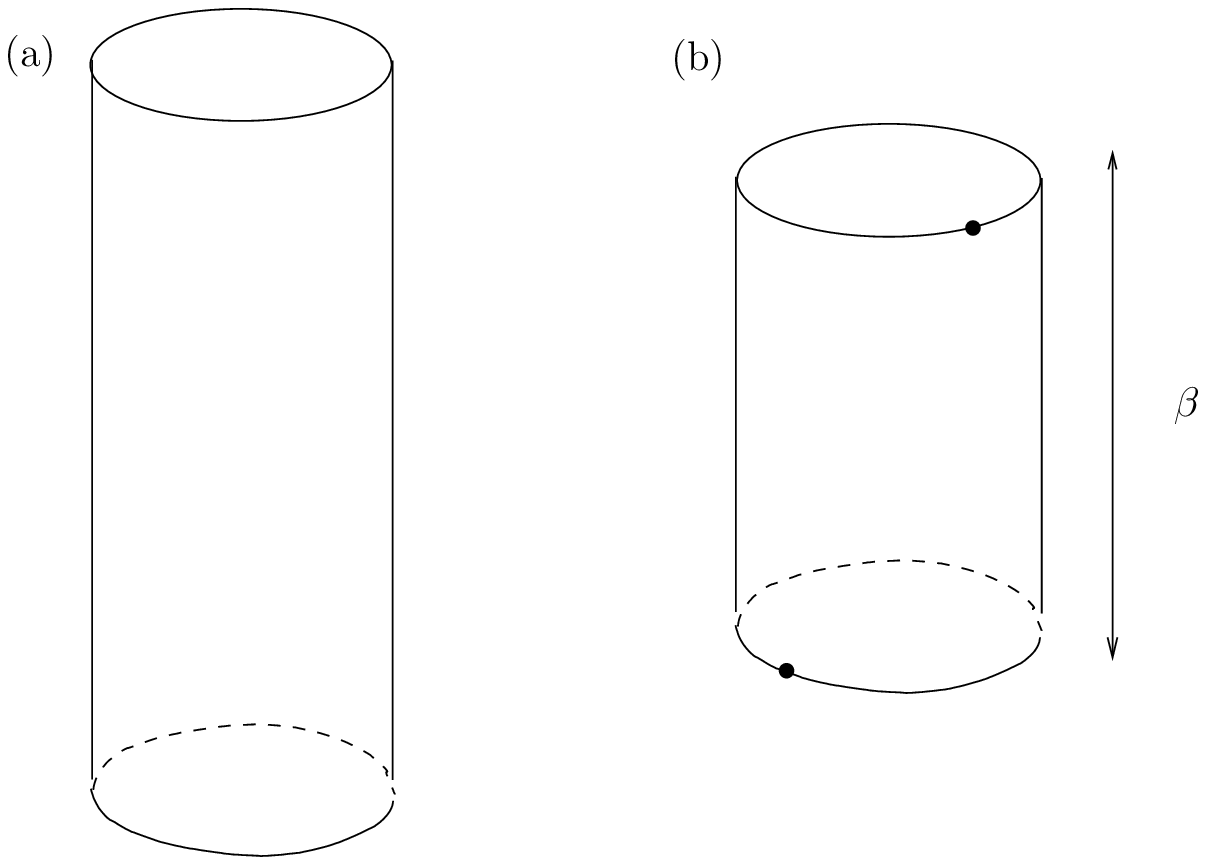}{9cm}{7cm}

The simplest such manifold is $M_{0,1}$, which we will now
describe in more detail. If we identify the real one-parameter
subgroup ${\rm diag}(e^b,e^{-b})$ of $SL(2,\C)$ as the group of
time translations, the ${\rm AdS}_3$ metric \adsmet\ can be put in
the form \eqn\hdr{ds^2=\cosh^{2}r \,\,dt^2 + dr^2+\sinh^{2}r\,\,
d\phi^2,} with $-\infty<t<\infty$, $0\leq r<\infty$, and
$0\leq\phi\leq 2 \pi$. Described in this coordinate system is the
subset $\AdS_3'\subset \AdS_3$ on which $\Gamma $ acts nicely;
its topology is $D\times \R$, where $D$ is a two-dimensional open
disc parameterized by $r$ and $\phi$. Conformal infinity is at
$r=\infty$.  (See Fig. 1.) The element ${\rm diag}(e^b,e^{-b})$
acts by $t\to t+b$. The group of spatial rotations is the
one-parameter group ${\rm diag}(e^{i\theta},e^{-i\theta})$, acting
by $\phi\to\phi+\theta$.

The group element $W$ therefore generates a combined
time-translation and spatial rotation.  To explicitly divide by
$W$, we ``cut'' $\AdS_3'$ at times $t=0$ and $t=2\pi\, {\rm
Im}\,\tau$. Then we glue together the top and bottom of the region
$0\leq t\leq 2\pi\, {\rm Im}\,\tau$ after making a spatial
rotation by an angle $2\pi\, {\rm Re}\,\tau$.  This is sketched in
Fig. 1b. The resulting spacetime $M_{0,1}$ is topologically
$D\times S^1$, so its fundamental group is indeed $\Z$.

The path integral in this spacetime has a simple semiclassical
meaning, since it may be interpreted in terms of Hamiltonian time
evolution.  A state is prepared at time zero and propagates a
distance $\beta=2\pi\, {\rm Im}\,\tau$ forward in Euclidean time.
In this process, the state vector is multiplied by the time
evolution operator $\exp(-\beta H)$, where $H$ is the Hamiltonian.
Then, after a spatial rotation by an angle $\theta=2\pi\, {\rm
Re}\,\tau$, which acts on the state by $\exp(-i\theta J)$, we glue
the top and bottom of the figure, which results in taking the
inner product of the final state with the initial state. The whole
operation gives the trace $\Tr\,\exp(-\beta H-i\theta J)$ defined
in the Hilbert space of perturbative fluctuations around $\AdS_3$.
This Hamiltonian interpretation of the path integral of $M_{0,1}$
will be the basis for evaluating it in Sec. 2.2.

The other manifolds $M_{c,d}$ are obtained from $M_{0,1}$ by
modular transformations, that is, by diffeomorphisms that act
non-trivially on the homology of $\Sigma$.  This fact
will allow us to evaluate their contributions to the path
integral.  These other manifolds may be thought of as Euclidean black holes.

For example, in the $(t,\phi)$ coordinates introduced above,
$M_{0,1}$ involves the identifications \eqn\phit{\phi + i t \sim
\phi + i t + 2\pi \sim \phi + it + 2\pi \tau.} From the metric
\hdr, note that the $\phi$ circle is contractible in the bulk,
since the coefficient of $d\phi^{2}$ becomes zero at the origin
$r=0$. The manifold $M_{1,0}$ is described by the same coordinate
system \hdr, but with new identifications \eqn\phitt{\phi + i t
\sim \phi + i t + 2\pi \sim \phi + it-{2\pi\over \tau}.} These
identifications \phitt\ may be written in the form of \phit\ by
taking $ {\phi} + i {t} \to {1\over \tau} \left(\phi + i
t\right)$. After this transformation, the coordinate that is
contractible in the bulk now involves a combination of $ \phi$ and
$t$.

If we take $\tau$ to be imaginary, the scaling by $1/\tau$ has the
effect of exchanging (and rescaling) $\phi$ and $t$.  $M_{1,0}$ is
hence a Euclidean black hole, in which the ``time'' circle is
contractible rather than the ``space'' circle. If we rotate to
Lorentzian signature by $t\to i t$, then the locus $r=0$ where the
coefficient of $dt^2$ vanishes is the horizon of the black hole.
In fact, $M_{1,0}$ is just the Euclidean version of the BTZ black
hole \carliptwo.  The more general manifolds $M_{c,d}$ are
often referred to as the $SL(2,\Z)$ family of black holes.

\bigskip\noindent{\it Orbifolds}

Finally, let us consider the extension in which $\Gamma$ has a
second generator given by \enby.  We may assume that $n=2m$ is
even, since the element ${\rm diag}(-1,-1)$ (the non-trivial
element of the center of $SL(2,\C)$)  acts trivially on $\CP^1$.
We also  may as well assume that $m>1$, since the case $m=1$ leads
to nothing new.

What we get when $\Gamma$ has an additional generator with $m>1$
is simply a three-dimensional space of the form $M_{c,d}/\Z_m$.
The conformal boundary is still a Riemann surface of genus 1, and
by changing $q$ we can adjust its modular parameter as we wish.
However, for $m>1$, the group element \enby\ acts on $\AdS_3'$
with fixed points, meaning that $M_{c,d}/\Z_m$ has orbifold
singularities. The fixed points are of codimension 2 and the
singularities look locally like $\R^2/\Z_m$, where $\Z_m$ acts as
a rotation by an angle $2\pi/m$.  This produces a deficit angle
\eqn\deficit{\theta=2\pi(1-1/m).}  The picture is sketched in Fig.
2.

\fig{The black line represents an orbifold singularity in the
interior of a spacetime that at infinity looks like $\AdS_3$.}
{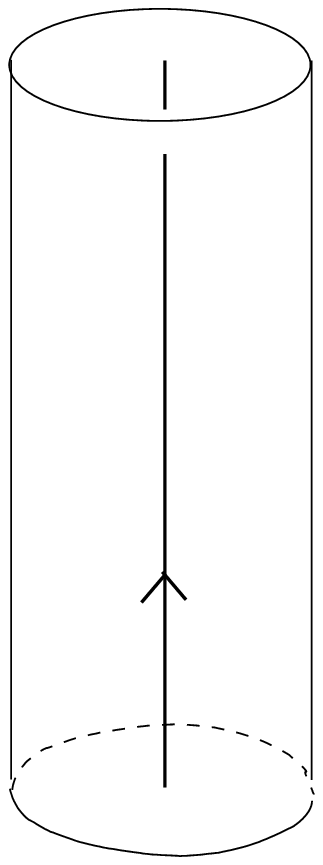}{4cm}{7cm}

At the classical level, the physical meaning in three-dimensional
gravity of  a codimension two singularity  characterized by a
deficit angle is usually that it represents the orbit of a massive
particle.  The mass of the particle is related to the deficit
angle \jdt.   The particular values of the deficit angle in eqn.
\deficit\ are special because they are related to orbifolds.

 Of course, there are many
consistent  theories of three-dimensional quantum gravity plus
matter, obtained from various widely studied string theory and
$M$-theory constructions.  They each have massive particles of
various sorts, the precise masses and spins being model-dependent.
It would be hopeless to try to completely solve such a general
theory.

The question of interest in the present paper is whether there
exists a theory that is in some sense minimal and  solvable. Our
hypothesis is that the minimal theory should be described by
smooth geometries without orbifold or deficit angle singularities.
 Pragmatically, it appears
difficult to avoid the problem found in Sec. 3 simply by including
a small set of such singularities.  In any event, were we to allow
singularities, we would not know which ones to allow.

In the dual CFT, massive particles correspond to primary operators
of positive dimension.  Any two-dimensional CFT -- and they are
abundant, of course -- can be interpreted as a dual $\AdS_3$
theory.  If this dual theory has a macroscopic, semiclassical
interpretation, it can be interpreted in terms of $\AdS_3$ with
particles and black holes.  From the holographic point of view, it
is difficult to understand which CFTs have a semiclassical
interpretation, but it is clear that, whether or not
there is what one might call a minimal theory, there are many
consistent theories with massive particles.

\subsec{Evaluation Of The Partition Function}

We write $Z_{c,d}(\tau)$ for the contribution to the partition
function of the manifold $M_{c,d}$.  Because the manifolds
$M_{c,d}$ are all diffeomorphic to each other, the functions
$Z_{c,d}(\tau)$ can all be expressed in terms of any one of them,
say $Z_{0,1}(\tau)$, by a modular transformation.  The formula is
simply \eqn\bufo{Z_{c,d}(\tau)=Z_{0,1}((a\tau+b)/(c\tau+d)),}
where $a$ and $b$ are any integers such that $ad-bc=1$.  The
partition function, or rather the sum of known contributions to
it, is
\eqn\telcox{Z(\tau)=\sum_{c,d}Z_{c,d}(\tau)=\sum_{c,d}Z_{0,1}((a\tau+b)/(c\tau+d)).}  The summation here is over all integers $c$ and $d$ which are relatively prime and have $c\ge0$.

This formula shows that the key point is to evaluate
$Z_{0,1}(\tau)$.  We recall that this contribution is simply
$\Tr\,\exp(-\beta H-i\theta J)$, computed in the Hilbert space
that describes small fluctuations about $\AdS_3$ (as opposed to
black holes).  If we know the eigenvalues of the commuting
operators $H$ and $J$ in the Hilbert space of small fluctuations,
then we can compute the trace.

In the most naive semiclassical approximation, $Z_{0,1}(\tau)$ is
just $\exp(-I)$, where $I$ is the classical action. In computing
this action, one can not just evaluate the action \harrigo\ for
the solution \hdr; such a computation would give an infinite
answer, coming from the boundary at $r\to\infty$.  The full action
includes the Gibbons-Hawking boundary term, which has the opposite
sign of the Einstein-Hilbert term \harrigo. This extra term
removes the divergence, and one arrives at a finite (negative)
answer for the action of $M_{0,1}$ \carliptwo: \eqn\gopher{I =
-4\pi k {\rm Im}\,\tau} where $k=\ell/16 G$.  Therefore, in this
approximation, we have \eqn\belox{Z_{0,1}(\tau)\simeq |\bar q
q|^{-k}.}

The result \belox\ has a simple interpretation, in terms of facts
explained in \strom. Three-dimensional pure gravity with the
Einstein-Hilbert action \harrigo\ is dual to a conformal field
theory with central charge $c_L=c_R=3\ell/2 G=24k$.   The formula
$c_L=c_R=3\ell/2G$ is actually the semiclassical approximation of
\brownhen.  Depending on how the theory is regularized, there may
be quantum corrections to this formula, but they preserve
$c_L=c_R$ because the Einstein-Hilbert theory is parity symmetric.
(By adding to the action a gravitational Chern-Simons term, we
could generalize to $c_L\not= c_R$.  We omit this here.) For our
purposes, we simply parametrize the theory in terms of
$k=c_L/24=c_R/24$.  Since $c_L$ and $c_R$ are physical observables
(defined in terms of the two-point function of the stress tensor
in the boundary CFT, or an equivalent bulk computation), the
theory when parametrized in this way does not depend on any choice
of formalism.  (By contrast, if we describe the results in terms
of the microscopic variable $\ell/G$, then it is also necessary to
describe the regularization.) It is conjectured in \WittenKT\ that
pure three-dimensional quantum gravity exists, if at all, only for
integer $k$.  However, for our present purposes, we do not need to
know if this is correct.

Let $L_0$ and $\tilde L_0$ be the Hamiltonians for left- and
right-moving modes of the CFT. They are related to what we have
called $H$ and $J$ by \eqn\morego{\eqalign{H&=L_0+\tilde L_0\cr
         J & = L_0-\tilde L_0.\cr}}
The CFT ground state has $L_0=-c_L/24$, $\tilde L_0=-c_R/24$, or
in the present context $L_0=\tilde L_0=-k$.  Equivalently, this
state has $H=-2k$, $J=0$.  Its contribution to
$\Tr\,\exp\left(-2\pi({\rm Im}\,\tau)H+2\pi i{({\rm Re}\,\tau})
J\right)$ is $\exp(4\pi k {\rm Im}\,\tau)=|\bar q q|^{-k}$, as in
eqn. \belox.

Thus, although seemingly only a naive approximation, eqn. \belox\
actually gives the exact result for the contribution of the ground
state to the partition function (if the theory is parametrized in
terms of the central charges).  As a result, it is also exact for
${\rm Im}\,\tau\to\infty$.

To get a complete answer, we also need to know the energies of
excited states.  In more than $2+1$ dimensions, the trace would
receive contributions from a gas of gravitons as well as from
other particles, if present.  Their energies would in general
receive complicated perturbative (and perhaps nonperturbative)
corrections.  Hence, although the function $Z_{0,1}(\tau)$ has a
natural analog in any dimension, above $2+1$ dimensions one would
not expect to be able to compute it precisely.

\def\OMEGA{|\Omega\rangle}
 In $2+1$ dimensions, there are no gravitational waves. Naively,
there are no perturbative excitations at all above the $\AdS_3$
vacuum, and hence one might at first expect the formula \belox\ to
be exact.  However, this is mistaken, because of the insight of
Brown and Henneaux \brownhen.  There must at least be states that
correspond to Virasoro descendants of the identity, or in other
words states obtained by repeatedly acting on the CFT vacuum
$\OMEGA$ with the stress tensor.  If $L_n$ and $\tilde L_n$ are
the left- and right-moving modes of the Virasoro algebra, then a
general such state is \eqn\tryto{\prod_{n=2}^\infty
L_{-n}^{u_n}\prod_{m=2}^\infty \tilde L_{-m}^{v_m}\OMEGA,} with
non-negative integers $u_n,v_m$.  (For a state of finite energy,
almost all of these integers must vanish. The products begin with
$n,m=2$ since $L_{-1}$ and $\tilde L_{-1}$ annihilate the CFT
ground state.) A state of this form is an eigenstate of $L_0$ and
$\tilde L_0$ with $L_0=-k+\sum_{n=2}^\infty n u_n$, $\tilde
L_0=-k+\sum_{m=2}^\infty m v_m$.  The contribution of these states
to the partition function is then \eqn\zelox{Z_{0,1}(\tau)= |\bar
q q|^{-k}{1\over \prod_{n=2}^\infty|1-q^n|^2}.} It is convenient
to introduce the Dedekind $\eta$ function, defined by
\eqn\dedfn{\eta(\tau)=q^{1/24}\prod_{n=1}^\infty (1-q^n).} Eqn.
\zelox\ can then be rewritten \eqn\felox{Z_{0,1}(\tau)={1\over
|\eta(\tau)|^2}|\bar q q|^{-(k-1/24)}|1-q|^2.} This formula will
be useful in Sec. 3 because $({\rm Im}\,\tau)^{1/2}|\eta(\tau)|^2$
is modular-invariant.

We will show shortly how the Virasoro descendants of eqn. \tryto\
arise in an analysis along the lines of Brown and Henneaux.  The
argument will also show that the formula  \zelox\ is exact to all
order of perturbation theory.  As for whether there are
non-perturbative corrections to eqn. \zelox, it is not clear
whether this question is well-defined, since it may be impossible
to separate the question of unknown nonperturbative corrections to
$Z_{0,1}$ from the more general question of unknown
nonperturbative corrections to the exact partition function $Z$.
At any rate, it will be easier to discuss what nonperturbative
corrections would mean after explaining why eqn. \zelox\ agrees
with perturbation theory.

Before presenting this argument, let us discuss what the answer
means from the general point of view of quantum mechanical
perturbation theory. Usually, one defines an effective action
$I_{\rm eff}$ such that the partition function is $Z=\exp(-kI_{\rm
eff})$.  $I_{\rm eff}$ is equal to the classical action $I$ plus
contributions generated from $r$-loop diagrams: \eqn\tefro{I_{\rm
eff}=I+\sum_{r=1}^\infty k^{-r}I_r.} Here $I_r$ is a function
generated from Feynman diagrams with $r$ loops.  In a general
theory, the expansion \tefro\ is only asymptotic in $1/k$; there
may be further contributions that are non-perturbatively small for
$k\to \infty$. However, in the present context, the formula
\zelox\ implies that (for perturbation theory around the manifold
$M_{0,1}$) there are no corrections beyond one-loop order; $I_{\rm
eff}$ is simply the classical action plus a one-loop correction.
In other words, the formula implies that if $Z_{0,1}$ is directly
computed in perturbation theory, then the perturbation series
terminates with the one-loop term.  (Such a direct computation is
also briefly discussed below.)

In a sense, this should not come as a surprise.  The gauge theory
interpretation of three-dimensional gravity
\refs{\townsend,\witten} (or even more naively the absence of any
local excitations) suggests that in some sense it is an integrable
system.  It is often true for quantum integrable systems that
properly chosen quantities are one-loop exact.

\bigskip\noindent{\it Derivation Of The Formula}

In most of physics, the quantum Hilbert space of a theory is
constructed by quantizing an appropriate classical phase space.
This framework may ultimately be inadequate for quantum gravity,
but it is certainly adequate for constructing the perturbative
Hilbert space that is needed to compute $Z_{0,1}$ in perturbation
theory.

The phase space of any physical theory is simply the space of its
classical solutions.  In $2+1$-dimensional gravity with AdS
boundary conditions at infinity, naively speaking the phase space
relevant to perturbation theory consists of only a single point,
since any solution that is close to the $\AdS_3$ solution (close
enough to exclude black holes) actually is diffeomorphic to
$\AdS_3$.

However, we must be more careful here.  In General Relativity, one
should divide only by diffeomorphisms that approach the identity
fast enough at infinity. After doing so, one constructs the
classical phase space $\cal M$; it parametrizes classical
solutions that obey the boundary conditions modulo diffeomorphisms
that vanish fast enough at infinity. One is then left with an
action of a group $G$ that consists of those diffeomorphisms that
preserve the boundary conditions, modulo those that vanish fast
enough at infinity that they are required to act trivially on
physical states.

\def\DIFF{\widehat{\rm diff}}
\def\diff{{\rm diff}}
In General Relativity in $3+1$ or more dimensions, in a spacetime
that is asymptotically Lorentzian, the group $G$ is the Poincar\'e
group.  This is why quantum gravity, in an asymptotically
Lorentzian spacetime, is Poincar\'e invariant. In
$2+1$-dimensional gravity with negative cosmological constant, the
obvious analog of this answer would be $SL(2,\R)\times SL(2,\R)$
(or a group locally isomorphic to that one), the group of
symmetries of $\AdS_3$. However, Brown and Henneaux showed
\brownhen\ that the actual answer turns out to be the
infinite-dimensional group $G=\diff\,S^1\times \diff\,S^1$ (which
contains $SL(2,\R)\times SL(2,\R)$ as a subgroup). By computing
Poisson brackets, they also showed that the quantum symmetry group
will be not $G$ but a central extension with (in a semiclassical
approximation) $c_L=c_R=3\ell/2G$.  We denote this central
extension as $\DIFF\,S^1\times \DIFF\, S^1$.

\def\M{{\cal M}}
We will carry the  analysis of \brownhen\ just slightly farther to
describe the phase space $\cal M$ and the resulting energy levels.
  First of all,
$\cal M$ is a homogeneous space for $G$, since if we divide by all
diffeomorphisms, then the classical solutions that obey the
boundary conditions (and are close enough to $\AdS_3$) are
equivalent to $\AdS_3$.  So $\M$ is  $G/H$, where $H$ is some
subgroup of $G$.  In fact, $H$ is the subgroup of $G$ that leaves
fixed a given point on $\M$.  Differently put, the symmetry group
of a point on $\M$ is isomorphic to $H$. In the present context,
we simply pick the point on $\M$ corresponding to $\AdS_3$ and
observe that its symmetry group is $SL(2,\R)\times SL(2,\R)$.  So
this is $H$, and the phase space is $\M=(\diff\,S^1\times
\diff\,S^1)/(SL(2,\R)\times SL(2,\R))=(\diff\,S^1/SL(2,\R))^2$.

In general, the quantization of a homogeneous
space $G/H$ should be expected to give a Hilbert space that is an
irreducible representation $R$ of $G$ (or more generally of a
central extension of $G$).  Moreover, as $G/H$ contains an
$H$-invariant point, the representation $R$ will contain a vector
that is an eigenvector for the action of $H$ (and hence is
$H$-invariant if $H$ is a simple non-abelian group).

In the present problem, a representation of the Virasoro group
$\DIFF\,S^1$ in which $L_0$ is bounded below and in which there is
an $SL(2,\R)$-invariant vector is uniquely determined once the
central charge $c=24k$ is given.  It is the ``vacuum''
representation, containing a vector $\OMEGA$ with
$(L_n+k\delta_{n,0})\OMEGA=0$, $n\geq -1$.  This representation is
spanned by states of the form \eqn\mitro{\prod_{n=2}^\infty
L_{-n}^{a_n}\OMEGA, } with energy
\eqn\itro{\epsilon=-k+\sum_{n=2}^\infty n a_n.} Quantizing
$(\diff\,S^1/SL(2,\R))^2$ gives a tensor product of two such
representations for the two factors of $G=(\diff\,S^1)^2$, and in
this way we arrive at the spectrum claimed in eqn. \tryto, and
hence at the formula \zelox\ for the partition function.

Quantization of homogeneous spaces of $\diff\,S^1$ is described in
\coadj.  The fact that quantization of the quotient
$\diff\,S^1/SL(2,\R)$ gives the vacuum representation of
$\DIFF\,S^1$ is explained from several points of view in eqns.
(170), (172), and (174) of that paper.

The states that we have described can be regarded as boundary
excitations, supported near the boundary of $\AdS_3$, since $\M$
collapses to a point if we are allowed to make general changes of
coordinate near the boundary.  We call these boundary excitations
the Brown-Henneaux (BH) states, since they are so closely related
to the analysis in \brownhen.

\bigskip\noindent{\it Corrections?}

Now let us ask to what extent there may be corrections to the
spectrum just described.

First we consider the question of quantum corrections to the
formula \itro\ for this family of states.  We can dispose of this
question immediately: this particular family of states transforms
in an irreducible representation of the symmetry group, and this
representation has no possible deformations once the central
charge is fixed.

So there are no quantum corrections to the energies of the states
just described. Could there be additional states contributing to
$Z_{0,1}$?

In general, in three-dimensional gravity, there may be additional
states of various kinds, corresponding for instance to massive or
even massless particles.  What do we expect in ``pure''
three-dimensional gravity, if it exists?   It is important to
distinguish two length scales: the $\AdS_3$ length scale $\ell$,
and the Planck length $G$.  In the semiclassical regime of
$k\to\infty$, we have $G<<\ell$.

In eqn. \itro, energies are measured in units of $1/\ell$.  A
reasonable minimum condition for anything that one might call a
theory of pure gravity is that the energy of any state that is not
part of the BH spectrum  goes to infinity in the semiclassical 
limit $k\to\infty$ (when measured relative to the energy of the
ground state). States that are not part of the BH spectrum and
whose excitation energies remain fixed for $k\to\infty$ would be
interpreted in terms of non-gravitational fields propagating in
$\AdS_3$.

As for what sort of states might have energies that go to infinity
for $k\to \infty$, we certainly expect BTZ black holes, with
energy bounded below by $k$.  As for what else there may be, the
sky is the limit in terms of conceivable speculations. If so
inclined, one can postulate solitons with excitation energies
proportional to $k$, states similar to $D$-branes with energies
proportional to $k^{1/2}$, etc. Orbifold singularities with fixed
deficit angle, as described in relation to eqn.
\deficit, would also have excitation energy of order $k$.

Our point of view is that the question of whether there are
nonperturbative corrections to $Z_{0,1}$ from states whose
excitation energy  diverges as $k\to\infty$ is hard to separate
from the more general question of unknown nonperturbative
corrections to the exact partition function $Z$. Since the
perturbative evaluation of $Z_{0,1}$ gives a convergent and
physically sensible result (which is even one-loop exact), we may
as well regard hypothetical contributions in which $M_{0,1}$ is
enriched with a soliton, a $D$-brane, an orbifold singularity, or
some other unknown type of excitation as representing different,
presently unknown sectors of the path integral. In this sense, the
formula for $Z_{0,1}$ is exact.

\bigskip\noindent{\it Comparison To Perturbation Theory}

Since we claim that the formula for $Z_{0,1}$ is one-loop exact,
the question arises of why not to simply calculate it by
evaluating the relevant one-loop determinants.

We know of two efforts to do so.  In \carliptwo, a formula much
simpler than our result \zelox\ is claimed.  The analysis relies
on the relation of the relevant product of determinants to
Ray-Singer and Reidemeister torsion \raysinger.  For an
irreducible flat connection on  a compact manifold without
boundary, the torsion is simply a number, but for a manifold with
boundary (such as $M_{0,1}$ effectively is), the torsion must be
understood as measure on a certain moduli space associated with
the boundary. A rather subtle treatment of this is needed, we
suspect, to compute the one-loop correction using its
interpretation via torsion.

On the other hand, in \italian, the one-loop correction was
studied by expressing the determinants in terms of the appropriate
heat kernels, which were evaluated using a method of images
(analogous to the derivation of the Selberg trace formula).  We
believe that this method is conceptually completely correct.  The
claimed result is qualitatively similar to our formula \zelox\
(for example, it has an expansion in integer powers of $q$ and
$\bar q$), and we hope that it will prove possible to resolve any
discrepancies between the result in \italian\ and the one claimed
here.

As for why it is much simpler to compute the one-loop correction
via the Hamiltonian route that we have followed, this should not
really come as a surprise.  In general, path integrals on a
product $S^1\times Y$ are often most easily evaluated by
constructing an appropriate Hilbert space in quantization on $Y$
and then taking a trace.\foot{An excellent example of this
statement is given by Chern-Simons gauge theory with a compact
gauge group for the case that $Y$ is a compact Riemann surface
without boundary.  The partition function on $S^1 \times Y$ is an
integer, the dimension of the physical Hilbert space ${\cal H}$
associated with $Y$.  It is not too hard to describe ${\cal H}$,
compute its dimension, and thereby learn the value of the path
integral on $S^1\times Y$.  But direct evaluation of this path
integral by Lorentz-covariant methods is difficult, even in
perturbation theory.} That is especially likely to be true in the
present situation, in which the relevant excitations are rather
subtle boundary excitations, already known but difficult to
rediscover.

\subsec{The General Form Of The Partition Function}

We conclude with an analysis of the following question: in
general, what sort of function can be written as $\Tr\,\exp(-\beta
H)$, where $H$ is a hermitian operator on a Hilbert space ${\cal
H}$?  We assume that $H$ is constrained so that $\Tr\,\exp(-\beta
H)$ is convergent whenever ${\rm Re}\,\beta>0$.  (According to
standard assumptions about pure three-dimensional gravity, this
condition is satisfied, but we discuss in Sec. 4.1 one way that it
might fail.)

For $\Tr\,\exp(-\beta H)$ to be convergent, the number $n_E$ of
states of energy  less  than $E$ must be finite for each $E$.
Indeed, the trace is bounded below by $n_E \exp(-\beta E)$.
Finiteness of $n_E$ for all $E$ implies in particular that $H$
must have a discrete spectrum.

Let $E_*$ be any value of the energy and let $E_1,\dots, E_n$ be
the energy eigenvalues that are no greater than $E_*$.  Their
contribution to the partition function is $\sum_{i=1}^n\exp(-\beta
E_i)$, and the full partition function is therefore
\eqn\begox{\Tr\,\exp(-\beta H)=\sum_{i=1}^n\exp(-\beta E_i)+{\cal
O}(\exp(-\beta E_*)).} Here ${\cal O}(\exp(-\beta E_*))$ is a
function that is bounded by a multiple of $\exp(-\beta E_*)$.

We have to be careful here with one point:  any finite sum of
states of energy greater than $E_*$ makes a contribution of the
form ${\cal O}(\exp(-\beta E_*))$, but in general an infinite set
of states with energy $\geq E_*$ may make a contribution that is
not bounded in this way.  However, our hypothesis that
$\Tr\,\exp(-\beta H)$ converges whenever ${\rm Re}\,\beta>0$
ensures that the number $n_E$ of states of energy less than $E$
grows with $E$ more slowly than any exponential. This is enough to
justify the error estimate in eqn. \begox.

Our problem is slightly more general.  We have a pair of commuting
hermitian operators $H$ and $J$, and we are computing the trace
$\Tr\,\exp(-\beta H-i\theta J)$.  However, the eigenvalues of $J$
are integers, so \eqn\kino{\Tr\,\exp(-\beta H-i\theta
J)=\sum_{j\in \Z}e^{-ij\theta}\,\Tr_j\,\exp(-\beta H),} where
$\Tr_j$ is a trace in the subspace ${\cal H}_j$ in which $J$ acts
with eigenvalue $j$.  The functions $\Tr_j\exp(-\beta H)$ should
be constrained exactly as in eqn. \begox.

In Sec. 3, we will find instead that the evaluation of known
contributions to $\Tr_0\exp(-\beta H)$ take the form just
described up to a certain energy $E_*$, beyond which this form
breaks down.  The result looks like
 \eqn\hofu{\sum_{i=1}^n\exp(-\beta E_i )+\exp(-\beta
E_*)(f+\sum_{s=1}^\infty f_s \beta^{-s})} where $f$ is a negative
integer, rather than being positive as it should be, and the
corrections decay as power laws, rather than exponentials. (The
series in \hofu\ may be only asymptotic.) Both the fact that $f$
is negative and the fact that there are power-law corrections
mean that this function cannot be interpreted as
$\Tr\,\exp(-\beta H)$.

Finally, let us compare  the  assertion that $\Tr\,\exp(-\beta H)$
can only converge if $H$ has a discrete spectrum  with standard
results about the thermodynamics of physically sensible systems
with continuous spectrum, such as a free gas of particles on the
real line.  What is usually computed in such a case is the free
energy per unit volume; the total free energy is infinite simply
because the volume is infinite.  To make the total free energy
finite, one can place the gas in a finite volume; then the
spectrum is discrete and $\Tr\,\exp(-\beta H)$ converges.  A gas of particles in Anti-de Sitter space will have a discrete spectrum and finite partition function $\Tr\, \exp(-\beta H)$ as well. This is because the motion of any finite energy particle in Anti-de Sitter space is restricted to a finite volume region in the interior of AdS.

\newsec{Computing the Sum over Geometries}

As explained in Secs. 1 and 2, the known contributions to the
partition function of pure gravity in a spacetime asymptotic to
$\AdS_3$ come from smooth geometries $M_{c,d}$, where $c$ and $d$
are a pair of relatively prime integers (with a pair $c,d$
identified with $-c,-d$). Their contribution to the partition
function, including the contribution from the Brown-Henneaux
excitations, is \eqn\zsum{Z(\tau) = \sum_{c,d} Z_{0,1}(\gamma
\tau),} where \eqn\gammais{\gamma \tau = {a\tau+b\over
c\tau+d},~~~~~~~~~\gamma = \left(a\ b \atop c\ d\right) \in
SL(2,{\Z})} and \eqn\asd{Z_{0,1}(\tau) = \left|q^{-k}
\prod_{n=2}^{\infty}(1-q^{n})^{-1}\right|^{2} = {|\bar q
q|^{-k+1/24}|1-q|^{2}\over |\eta(\tau)|^{2}} .} The summation in \zsum\ is over all relatively prime $c$ and $d$ with $c\ge 0$.
Since
$Z_{0,1}(\tau)$ is invariant under $\tau\to\tau+1$,  the summand
in \zsum\ is independent of the choice of $a$ and $b$ in \gammais.
This sum over $c$ and $d$ in  \zsum\ should be thought of as a sum
over the coset $PSL(2,\Z)/\Z$, where $\Z$ is the subgroup of
$PSL(2,\Z)=SL(2,\Z)/\{\pm 1\}$ that acts by $\tau\to\tau+n$,
$n\in\Z$.  Given any function of $\tau$, such as $Z_{0,1}(\tau)$,
that is invariant under $\tau\to\tau+1$, one may form a sum such
as \zsum, known as a Poincar\'e series.

The function $\sqrt{{\rm Im}\, \tau}|\eta(\tau)|^2$ is
modular-invariant.  We can therefore write $Z(\tau)$ as a much
simpler-looking Poincar\'e series, \eqn\zzsum{Z(\tau)={1\over
\sqrt{{\rm Im}\, \tau}|\eta(\tau)|^2}\sum_{c,d}\left.\left(
\sqrt{{\rm Im}\, \tau}|\bar q
q|^{-k+1/24}|1-q|^{2}\right)\right|_\gamma,} where
$\left.(\dots)\right|_\gamma$ is the transform of an expression
$(\dots)$ by $\gamma$.  Writing out explicitly $|1-q|^2=1-q-\bar
q+\bar q q$, we see that we really need a sum of four Poincar\'e
series, each of the form
\eqn\pelfo{E(\tau;n,m)=\sum_{c,d}\left.\left(\sqrt{{\rm Im}\,
\tau} q^{-n}\bar q^{-m}\right)\right|_\gamma,} with $n-m$ equal to
0 or $\pm 1$. Precisely this sum, or rather its $s$-dependent
generalization introduced below, has been studied in Sec. 3.4 of
\Iwaniec, as was pointed out by P. Sarnak.\foot{Actually, the
generalization to arbitrary integer values of $n-m$ is considered
in \Iwaniec.  We would encounter the same generalization if we
modify the original Einstein-Hilbert action \harrigo\ to include
the gravitational Chern-Simons term.} If we set $\kappa=n+m$,
$\mu=m-n$, and use the fact that ${\rm Im}\,(\gamma \tau)={\rm
Im}\,\tau/|c\tau+d|^2$, then the basic Poincar\'e series can be
written \eqn\zkdef{E(\tau; \kappa,\mu)= \sqrt{{\rm
Im}\,\tau}\sum_{c,d} |c\tau+d|^{-1} \exp\left\{2\pi \kappa~ {\rm
Im}\ \gamma \tau + 2\pi i\mu~{\rm Re}\ \gamma \tau\right\}. } When
$\kappa = 0$ and $\mu=0$, this sum is a non-holomorphic
Eiseinstein series of weight $1/2$.  Sometimes we omit $\tau$ and
write just $E(\kappa,\mu)$. In terms of this function, the
partition function is \eqn\zfour{ \eqalign{Z(\tau) ={1\over
\sqrt{{\rm Im}\,\tau} |\eta(\tau)|^2}& \bigl(E(2k-1/12,0) +
E(2k+2-1/12,0)\bigr.\cr& - \bigl. E(2k+1-1/12,1) -
E(2k+1-1/12,-1)\bigl).\cr}}

\subsec{The Regularized Sum}

The sums \zkdef\ that define the Poincar\'e series we need are
divergent.  (Such divergences have been encountered before in
similar sums related to three-dimensional gravity
\refs{\farey,\rabadan,\ManschotZB}.)  Using \eqn\asd{ {a \tau + b
\over c \tau + d} = {a\over c} -{1\over c(c\tau +d)}, } one may
show that if $\tau=x+iy$ then \eqn\basd{\eqalign{{\rm
Im}\,(\gamma\tau)&={y\over (cx+d)^2+c^2y^2}\cr {\rm
Re}\,(\gamma\tau)&={a\over c}-{cx+d\over c((cx+d)^2+c^2y^2)}.\cr}}
For $\mu=0$, the exponential factor in the definition of $E$ is
\eqn\asd{ \exp\left\{2\pi \kappa~{\rm Im}\ \gamma\tau \right\}=
\exp\left\{ 2\pi \kappa~ {y \over (cx+d)^{2} +
c^{2}y^{2}}\right\},} and this goes to 1 for $c,d\to\infty$.
 So the first two terms in \zfour\
diverge linearly as $ \sum_{c,d}|c\tau+d |^{-1}$ at large $c$ and
$d$.  The other terms also diverge, though more slowly.

We claim that this divergence has a natural regularization. On the
upper half plane, which we call $\eusm H$, there is a natural
$SL(2,\R)$-invariant Laplacian:
\eqn\dofo{\Delta=-y^2\left({\partial^2\over
\partial x^2}+{\partial^2\over\partial y^2}\right).}
A short calculation shows that the function $y^{1/2}$ is an
eigenfunction of $\Delta$: $\Delta(y^{1/2})=(1/4)y^{1/2}$. Since
$\Delta$ is $SL(2,\R)$-invariant, the same is true of $({\rm
Im}\,\gamma\tau)^{1/2}$ for any $\gamma\in SL(2,\Z)$ (or even
$SL(2,\R)$): \eqn\blofo{\Delta \,\sqrt{\rm
Im\,(\gamma\tau)}={1\over 4}\sqrt{\rm Im\,(\gamma\tau)}.}

Using this, one may verify that although the Poincar\'e series for
$E(\tau;\kappa,\mu)$ is divergent, the corresponding series for
$(\Delta-1/4)E(\tau;\kappa,\mu)$ actually converges.  (This series
is obtained by acting termwise with $(\Delta-1/4)$ on the
Poincar\'e series for $E$.)  To see this, we just observe, using
\asd, that contributions in which derivatives appearing in
$\Delta$ act on the exponential factor in \zkdef\ get an extra
convergence factor of order $1/({\rm max}(c,d))^2$ and lead to a
convergent sum over $c$ and $d$. On the other hand, contributions
in which none of the derivatives act on the exponential actually
vanish, because of \blofo.

So $(\Delta-1/4)E$ requires no regularization.  Similarly, if we
set $F=\sqrt{{\rm Im}\,\tau}|\eta|^2$, then $(\Delta-1/4)(FZ)$
requires no regularization, where $Z$ is the partition function.
Since no regularization is required, we assume that the naive sum
gives $(\Delta-1/4)(FZ)$ correctly.

The Laplacian $\Delta$, acting on the Hilbert space of
square-integrable $SL(2,\Z)$-invariant functions on $\eusm H$, has
a continuous spectrum starting at $1/4$.  It also has a discrete
spectrum, but there are no discrete modes with eigenvalue $1/4$.
So the operator $\Delta-1/4$ is invertible acting on
square-integrable functions, and roughly speaking, we now want to
argue that once $(\Delta-1/4)(FZ)$ is known, $FZ$ is  uniquely
determined.

There is a subtlety in making this argument, because actually $Z$
and $FZ$ are not themselves square-integrable.  Rather, we expect
that $Z$ grows exponentially for $y\to\infty$, because the
$\AdS_3$ vacuum has negative energy $-2k$.    There are also
Brown-Henneaux excitations of this vacuum, again with negative
energy, and they contribute additional exponentially growing terms
for $y\to\infty$.  If we assume that all exponentially growing
contributions to $Z$ come from the known Brown-Henneaux states,
then this, together with a knowledge of $(\Delta-1/4)(FZ)$, is
enough to determine $Z$ uniquely.

But for our main application, we do not really need this assumption
(which appears to be not quite correct according to the analysis of
Sec. 3.3). The explicit calculation of Secs. 3.2 and 3.3 will give
us a function $Z$, compatible with the Poincar\'e series for
$(\Delta-1/4)(FZ)$, such that the fastest growing exponentials agree
with the Brown-Henneaux spectrum. The leading departure from the
contributions of that spectrum are determined at the end of Sec. 3.3
and take the form \eqn\butroo{\eqalign{ Z'={1\over
|\eta|^2}\Big(-6&+ {(\pi^{3}-6\pi)(11+24k)\over 9 \zeta(3)} y^{-1}
\cr &+{5(53\pi^{6}-882 \pi^{2}) + 528(\pi^{6}-90\pi^{2})k +
576(\pi^{6}-90\pi^{2})k^{2}\over 2430 \zeta(5)} y^{-2}+ {\cal
O}(y^{-3})\Big).\cr}}  This function cannot be written as a sum of
exponentials with positive integer coefficients, since the leading
coefficient is negative, and the corrections do not have the right
form.
 Any other candidate for
$Z$ would be obtained by adding a correction $\tilde Z$ such that
$(\Delta-1/4)(F\tilde Z)=0$.  But that equation is not obeyed by
any function $\tilde Z$ that would solve our problem; the equation
$(\Delta-1/4)(F\tilde Z)=0$ is not obeyed by $- Z'$ or by any
function that differs from it for $y\to\infty$ by a sum of
exponentials with positive integer coefficients. So the problem
that we will find is not affected by adding to the function $Z$
that we compute an additional contribution $\tilde Z$ that obeys
$(\Delta-1/4)(F\tilde Z)=0$.

\bigskip\noindent{\it More Convenient Alternative}

Although this argument is satisfying conceptually, it does not
give a convenient way to determine $Z$ in practice.  A much more
convenient method is to adapt $\zeta$-function regularization
\raysinger\  to this problem.

In the present context, the analog of $\zeta$-function
regularization is to replace the Poincar\'e series \pelfo\ by a
more general one depending on a parameter $s$:
\eqn\zelfo{E(\tau;s,n,m)=\sum_{c,d}\left.\left(({\rm Im}\,
\tau)^{s} q^{-n}\bar q^{-m}\right)\right|_\gamma.} Since $({\rm
Im}\,\gamma\tau)^s=y^s/|c\tau+d|^{2s}$, the series $E(\tau;s,n,m)$
converges for ${\rm Re}\,s>1$.  Our original problem concerns the
case $s=1/2$. As we will see shortly, $E(\tau;s,n,m)$, defined
initially for ${\rm Re}\,s>1$, can be analytically continued to
$s=1/2$ without any problem.  This analytic continuation gives a
natural way to define the original function $E(\tau;n,m)$ and
therefore the partition function $Z$. This turns out to be a very
practical and useful method to study $E(\tau;n,m)$.

The only problem with this approach is that  the physical meaning
of the parameter $s$ in three-dimensional gravity is unclear;
hence, it is not clear {\it a priori} that the analytic
continuation will give the right answer.  The argument involving
$(\Delta-1/4)(FZ)$ is clearer conceptually, because
$(\Delta-1/4)(FZ)$ is a physical observable and we simply use the
fact that the formal path integral expression for it converges.

However, it is not difficult to show that the two methods give the
same result.  For this, we observe that the function
$H(\tau;s,m,n)=(\Delta+s(s-1))E(\tau;s,m,n)$ can also be
represented by a Poincar\'e series, convergent when ${\rm
Re}\,s>0$.  For ${\rm Re}\,s>1$, where both Poincar\'e series
converge, the two functions $E$ and $H$, both defined by their
Poincar\'e series, obey $H=(\Delta+s(s-1))E$.  This automatically
remains true after analytic continuation to $s=1/2$.  So the
function $E(\tau;1/2,m,n)$  defined by analytic continuation from
${\rm Re}\,s>1$ has the property that
$(\Delta-1/4)E(\tau;1/2,m,n)$ is given by the obvious Poincar\'e
series.

In terms of $\kappa=m+n$, $\mu=m-n$, the regularized Poincar\'e
series is
 \eqn\poincare{ E(s,\kappa,\mu) =
 \sum_{c,d} {
y^{s} \over |c\tau + d|^{2s} } \exp\left\{2\pi \kappa~ {\rm Im}~
\gamma \tau + 2\pi i\mu~{\rm Re}~ \gamma \tau\right\}. } When
$\kappa=0$ and $\mu=0$, this sum is the non-holomorphic Eisenstein
series of weight $s$.

\subsec{Poisson Resummation}

Sec. 3.4 of \Iwaniec\ contains precisely what we need to analyze the
sum \poincare, make the analytic continuation, and determine if the
partition function $Z$ is physically sensible.

First, define $d=d'+nc$, where $n$ is an integer and $d'$ runs
from $0$ and $c-1$.  We may separate out the sum over $n$ in
\poincare\ to get
\eqn\bigsum{ E(s,\kappa,\mu) = y^{s}e^{2\pi
(\kappa y+i \mu x)} + \sum_{c>0}\,\sum_{d'\in\Z/c\Z}\,
\sum_{n\in{\Z}} f(c,d',n)} where\eqn\fis{ f(c,d',n) ={y^s\over
|c(\tau +n) + d'|^{2s}} \exp\left\{ {2\pi \kappa y \over |
c(\tau+n) +d'|^{2}} + 2\pi i \mu~ \left({a\over c} - {cx+d\over
c |c(\tau +n) +d'|^{2}}\right) \right\}. } The first term in eqn.
\bigsum\ comes from $c=0,d=1$.

The Poisson summation formula allows us to turn the sum over $n$
into a sum over a Fourier conjugate variable $\hat n$ \eqn\asd{
\sum_{n\in{\Z}} f(c,d',{ n}) = \sum_{\hat n\in{\Z}} {\hat
f}(c,d',{\hat n})} where ${\hat f}(c,d',{\hat n})$ is
the Fourier
transform \eqn\fint{\eqalign{ {\hat f} (c,d',{\hat n}) &=
\int_{-\infty}^{\infty} dn~ e^{2\pi i n {\hat n}} f(c,d',n)\cr
 &= \exp\left({2\pi i \left({\mu a-{\hat n}d'\over c}
- {\hat n } x\right)}\right) \int_{-\infty}^{\infty} d{t}~ e^{2\pi
i {\hat n} t} \left( {y\over c^{2}(t^{2}+y^{2})} \right)^{s}
\exp\left\{{ 2\pi (\kappa\,y -i\mu t)\over c^{2}(t^{2}+y^{2})}
\right\}.\cr}} We have written the integral in terms of a shifted
integration variable  $ t= n + x + {d'\over c}$. Upon Taylor
expanding the exponential that appears in the integral and
introducing $T=t/y$, we get \eqn\helpo{\eqalign{  {\hat f}
(c,d',{\hat n})&= \sum_{m=0}^{\infty}c^{-2(s+m)} e^{2\pi i
\left({\mu a-{\hat n}d'\over c} - {\hat n } x\right)}
{(2\pi)^{m}\over m!}\int_{-\infty}^{\infty} d{t}~ e^{2\pi i {\hat
n} t} \left( {y\over t^{2}+y^{2}} \right)^{m+s} \left(\kappa-i\mu
 {t\over y} \right)^{m}\cr &= \sum_{m=0}^{\infty}c^{-2(s+m)}
e^{2\pi i \left({\mu a-{\hat n}d'\over c} - {\hat n } x\right)}
{(2\pi)^{m}\over m!} y^{1-m-s}\int_{-\infty}^{\infty} d{T}
~e^{2\pi i {\hat n} T y} \left( {1+T^{2}} \right)^{-m-s}
\left(\kappa-i\mu  T \right)^{m}. }}

In \helpo, $c$ and $d'$ do not appear in the integrals but only in
the elementary prefactors, so we can study the sums over $c$ and
$d'$ explicitly. Note that, as $\mu$ is an integer, \helpo\ depends
on $a$ only modulo $c$.  The value of $a$ mod $c$ is determined by
$d'$, given that $ad'=1$ mod $c$. For a given $d'$, such an $a$
exists if and only if $d'$ lies in the set $(\Z/c\Z)^*$ of residue
classes mod $c$ that are invertible multiplicatively. So, dropping
the prime from $d'$, we may write the sum over that variable as
\eqn\klooster{ S(-{\hat n},\mu;c) = \sum_{d\in (\Z/c\Z)^*}
\exp\left\{2\pi i\left(-{\hat n} d + \mu d^{-1}\over c
\right)\right\}, } where $d^{-1}\in (\Z/c\Z)^*$ is the
multiplicative inverse of $d$. This sum is known as a Kloosterman
sum.

Rearranging the sums in \bigsum, we now have \eqn\bigsumm{
E(s,\kappa,\mu) = y^{s}e^{2\pi (\kappa y+i \mu x)} + \sum_{\hat
n}e^{-2\pi i {\hat n}x} E_{\hat n}(s,\kappa,\mu) } where
\eqn\eis{
E_{\hat n}(s,\kappa,\mu) = \sum_{m=0}^{\infty} I_{m,{\hat
n}}(s,\kappa,\mu)~ y^{1-m-s}~
\left(\sum_{c=1}^{\infty}c^{-2(m+s)}S(-{\hat n},\mu;c) \right) .}
Here we have defined the integral \eqn\idef{ I_{m,{\hat
n}}(s,\kappa,\mu) = {(2\pi)^{m}\over m!}\int_{-\infty}^\infty d{T}
~e^{2\pi i {\hat n} T y} \left( { 1+T^{2}} \right)^{-m-s}
\left(\kappa-i\mu T \right)^{m} .} Note that \eis\ is independent
of $x$, so that \bigsumm\ has the form of a Fourier expansion in
$x$ with Fourier coefficients $E_{\hat n}(s,\kappa,\mu)$ given by
\eis.  These Fourier coefficients are typically complicated
functions of $y$, since the integral \idef\ depends on $y$.

These Fourier coefficients are precisely what we want. In view of
\zfour, the Fourier expansion of the function $E$ with respect to
$x$ will give a similar expansion for the partition function $Z$.
The Fourier coefficients of $Z$ are the functions
$\Tr_j\,\exp(-\beta H)$, the partition function restricted to
states with angular momentum $J=j$.  These are the functions that
we want to understand.

We are almost ready to analyze  what happens when we continue the
formulas to $s=1/2$. For $m>0$, the integral in \idef\ is
convergent for ${\rm Re}\,s>0$, and likewise the sum
\eqn\dogood{\sum_{c=1}^{\infty}c^{-2(m+s)}S(-{\hat n},\mu;c)}
converges for ${\rm Re}\,s>0$.  A problem does occur for $m=0$,
since then neither the integral nor the sum is convergent at
$s=1/2$. We return to this shortly.

\subsec{The $\hat n=0$ Mode}

Let us first consider the Fourier mode which is constant in $x$,
i.e. the ${\hat n}=0$ term in \bigsumm. In this case the integral
\idef\ is independent of $y$, and may be evaluated explicitly. For
$\mu=0$, the result can be expressed in terms of $\Gamma$
functions \eqn\izero{ I_{m,0}(s,\kappa,0) =
\kappa^{m}{2^{m}\pi^{m+1/2}\Gamma(s+m-1/2)\over m! \Gamma(s+m)} ,}
while for $\mu=\pm1$, we require also hypergeometric functions
\eqn\ione{\eqalign{ I_{m,0}(s,\kappa,\pm1) = & \cos\left({m \pi
\over 2}\right){(2\pi)^{m}\Gamma\left({1+m\over 2}\right)
\Gamma\left({m-1\over2}+s\right)\over m!\Gamma(m+s)}
{}_{2}F_{1}\left({m-1\over2}+s,-{m\over
2}+s;{1\over2};\kappa^{2}\right) \cr &+m\kappa\sin\left({m \pi
\over 2}\right){(2\pi)^{m}\Gamma\left({m\over 2}\right)
\Gamma\left({m\over2}+s\right)\over m! \Gamma(m+s)}
{}_{2}F_{1}\left({1-m\over2},{m\over2}+s;{3\over2};\kappa^{2}\right)
.}} When $m=0$, this formula simplifies to \eqn\ionezero{
I_{0,0}(s,\kappa,\pm1) = \sqrt{\pi}{\Gamma(s-1/2)\over \Gamma(s)}.
}

The sum over $c$ may also be evaluated exactly (see Sec. 2.5 of
\Iwaniec).  For $\mu=0$ this evaluation involves the Kloosterman sum
$S(0,0;c)$; from the definition \klooster, one can see that
$S(0,0;c)$ is equal to the Euler totient function $\phi(c)$ (which
is defined as the number of positive integers less than $c$ that
are relatively prime to $c$). The sum over $c$ is a standard one
\eqn\szero{ \sum_{c=1}^{\infty}c^{-2(m+s)} S(0,0;c) =
\sum_{c>0}c^{-2(m+s)}\phi(c) = {\zeta(2(m+s)-1)\over\zeta(2(m+s))
}.} This formula can be obtained as follows. We start by noting a
basic property of the totient function: for any $n$, $\sum_{d|n}
\phi(d) = n$. In order to evaluate the sum $\sum_{c}
c^{-\sigma}\phi(c)$, let us multiply this sum by $\zeta(\sigma) =
\sum_{n} n^{-\sigma}$.  This gives \eqn\asd{\eqalign{
\zeta(\sigma) \sum_{c=1}^{\infty}c^{-\sigma}\phi(c) &=
\sum_{n,c=1}^{\infty}(nc)^{-\sigma}\phi(c)\cr &=
\sum_{m=1}^{\infty}m^{-\sigma} \sum_{c|m}\phi(c) \cr &=
\sum_{m=1}^{\infty}m^{1-\sigma} = \zeta(\sigma-1) .}} Setting
$\sigma=2(s+m)$ gives \szero.

For $\mu=\pm1$, the Kloosterman sum becomes a special case of what
is known as Ramanujan's sum (see Sec. 2.5 of \Iwaniec): \eqn\asd{
S({\hat n},0;c) = \sum_{d\in (\Z/c\Z)^*}e^{2\pi i d/c } = \mu(c).
} Here $\mu(c)$ is the M\"{o}bius function, which  is defined as
follows: $\mu(c)=0$ if $c$ is not square-free, while if $c$ is the
product of $k$ distinct prime numbers, then $\mu(c)=(-1)^k$. The
sum over $c$ is given by \eqn\sone{ \sum_{c=1}^{\infty}c^{-2(s+m)}
S(0,\pm1;c) = \sum_{c=1}^{\infty}c^{-2(s+m)} \mu(c) = {1\over
\zeta (2(s+m))}. } To prove this, one may use the basic property
of the M\"{o}bius function: for any $n$, $\sum_{d|n}
\mu(d)=\delta_{n,1}$.  To compute $\sum_{c} c^{-\sigma}\mu(c)$, we
multiply this sum by $\zeta(\sigma)$ to get \eqn\asd{\eqalign{
\zeta(\sigma) \sum_{c=1}^{\infty}c^{-\sigma}\mu(c) &=
\sum_{n,c=1}^{\infty} (cn)^{-\sigma}\mu(c)\cr &=
\sum_{m=1}^{\infty}m^{-\sigma} \sum_{c|m}\mu(c) \cr &=
\sum_{m=1}^{\infty}m^{-\sigma} \delta_{m,1}=1 .}} Setting
$\sigma=2(m+s)$ gives \sone.

Putting these formulae together gives an exact expression for the $x$
independent part of our Poincar\'e series \bigsum:
\eqn\poincaree{ E_{0}(s,\kappa,\mu) = \sum_{m=0}^{\infty} w_m(s,\kappa,\mu) y^{1-m-s}.
}
The constants $w_m(s,\kappa,\mu)$ in this expansion are independent
of $x$ and $y$, and are given explicitly by equations \izero,
\ione, \szero\ and \sone\  -- we will write them out in more detail for the cases of interest below.
The expansion \poincaree\ is one of our key results.  It is an explicit series expansion of our Poincar\'e series in powers of $y^{-1}$

Now we can study the behavior at $s=1/2$ of the delicate
contribution with $m=0$. For $\mu=0$, this term is finite because
the factor of $\Gamma(s-1/2)$ appearing in \izero\ and the factor
of $\zeta(2s)$ appearing in \szero\ both have simple poles at
$s=1/2$; as $s\to 1/2$, $\Gamma(s-{1/2})/\zeta(2s) \to 2$.
In fact, all of the complicated factors cancel to give
\eqn\eanszero{ E_0(1/2,\kappa,0) = - y^{1/2}+ {\cal O}(y^{-1/2}) .}
The coefficient of the leading term is negative because
$\zeta(0)=-1/2$. For $\mu=\pm1$, the $m=0$ term is finite because
the factors of $\Gamma(s-1/2)$ in \ionezero\ and $\zeta(2s)$ in
\sone\ both have simple poles at $s=1/2$.  We find
\eqn\eansone{
E_0(1/2,\kappa,\pm1) = 2 y^{1/2} + {\cal O}(y^{-1/2}) .}

Going back to \bigsumm, and writing simply $E(\kappa,\mu)$ for
$E(1/2,\kappa,\mu)$, we now have \eqn\bigrsum{E(\kappa,0)=
y^{1/2}\exp({2\pi \kappa y})- y^{1/2}+ {\cal O}(y^{-1/2}).}
Actually, we have not yet discussed the Fourier modes with $\hat
n\not=0$; however, the integrals \idef\ vanish exponentially for
$y\to\infty$ and so do not affect the assertion in eqn. \bigrsum.
Likewise, we have \eqn\bigrrsum{E(\kappa,\pm 1)= y^{1/2}\exp({2\pi
(\kappa y\pm i x)})+2 y^{1/2} + {\cal O}(y^{-1/2}).}

If we evaluate eqn. \zfour\ for the partition function $Z$ keeping
only the first terms in eqns. \bigrsum\ and \bigrrsum\ -- the
exponentially growing terms -- then the formula for $Z$ simply
reduces to $Z_{0,1}$, the contribution of the Brown-Henneaux
states.  What we have gained from all the work that we have done
is that we can now calculate corrections to $Z_{0,1}$.  The
leading corrections come from the corrections to the exponential
terms in \bigrsum\ and \bigrrsum.  Adding them up and evaluating
\zfour, we get \eqn\utro{Z=Z_{0,1}+{1\over
|\eta|^2}\left(-6+{\cal O}(y^{-1})\right).} Since
$1/|\eta|^2\sim |q|^{-1/12}$, $Z$ is governed by the
Brown-Henneaux spectrum up to energy $-1/12$ (slightly below the
classical black hole threshold, which is at zero energy).  The
number of states at that energy is not a positive integer, as one
would hope, but rather $-6$.

Moreover, the derivation shows that the corrections are given by a
power series in $1/y$, not a sum of exponentials.  As was
explained in Sec. 2.3, for $Z$ to have an interpretation as
$\Tr\,\exp(-\beta H)$, the corrections must be given by
exponentials.

Thus we have arrived at the main conclusion of this paper: the sum
of known contributions to the partition function of pure
three-dimensional gravity is not physically sensible.

It is illustrative to compute the next few terms in the expansion of the partition function \utro\ in powers of $y$.  This means taking $m\ne0$ in equations \izero, \ione, \szero\ and \sone.  Unlike the $m=0$ case, these expressions do not have poles at $s=1/2$, so the computation is straightforward.  Let us start with the $\mu=0$ case.
We find that the coefficients $w_{m}(s,\kappa,\mu)$ are given by
\eqn\wzero{
w_{m}(1/2,\kappa,0) = {2^{m}\pi^{m+1/2}\zeta(2m)\over m \Gamma(m+1/2)\zeta(2m+1)}\kappa^{m}
.}
So the next few terms in the Poincar\'e series \poincaree\ are
\eqn\asd{
E_{0}(1/2,\kappa,0) = -y^{1/2} +\left({2 \pi^{3}\over 3 \zeta(3)} \kappa\right) y^{-1/2} + \left({4\pi^{6}\over 135 \zeta(5)}\kappa^{2}\right)y^{-3/2} + {\cal O}(y^{-5/2})
}
To evaluate the $\mu=\pm1$ terms,
note that at $s=1/2$ the first two arguments of the relevant hypergeometric function appearing in \ione\ are integers.  This means that the formula \ione\ simplifies considerably at $s=1/2$ -- it is just a polynomial in $\kappa$.  It is
\eqn\ionesimp{
I_{m,0}({1\over 2},\kappa,\pm1) = {2 \pi^{m+1/2}\over m \Gamma(m+1/2)} T_{m}(\kappa)
}
where $T_{m}(\kappa)$ denotes a Chebyshev polynomial of the first kind.
So the coefficients appearing in \poincaree\ are
\eqn\wone{
w_{m}(1/2,\kappa,\pm1/2) = {2\pi^{m+1/2} \over m\Gamma(m+1/2)\zeta(2m+1)}T_{m}(\kappa) .
}
This allows us to write down the next few terms in the series:
\eqn\asd{
E_0(1/2,\kappa,\pm1) = 2 y^{1/2} + \left({4\pi\over\zeta(3)}\kappa\right)y^{-1/2} + \left({4\pi^{2}\over 3 \zeta(5)}(2\kappa^{2}-1)\right)y^{-3/2} + {\cal O}(y^{-5/2}) .
}

 Going back to \bigsumm, and writing $E(\kappa,\mu)$ for
$E(1/2,\kappa,\mu)$, we have \eqn\bigrsum{ E(\kappa,0)=
y^{1/2}\exp({2\pi \kappa y})- y^{1/2}+ \left({2 \pi^{3}\over 3
\zeta(3)} \kappa\right) y^{-1/2} + \left({4\pi^{6}\over 135
\zeta(5)}\kappa^{2}\right)y^{-3/2} + {\cal O}(y^{-5/2}) } and
\eqn\bigrrsum{E(\kappa,\pm 1)= y^{1/2}\exp({2\pi (\kappa y\pm i
x)})+2 y^{1/2} +
 \left({4\pi\over\zeta(3)}\kappa\right)y^{-1/2} + \left({4\pi^{2}\over 3 \zeta(5)}(2\kappa^{2}-1)\right)y^{-3/2} + {\cal O}(y^{-5/2})
.} Plugging this into \zfour\ gives the expansion of the partition
function \eqn\utroo{\eqalign{ Z=Z_{0,1}+{1\over |\eta|^2}\Big(-6&+
{(\pi^{3}-6\pi)(11+24k)\over 9 \zeta(3)} y^{-1} \cr
&+{5(53\pi^{6}-882 \pi^{2}) + 528(\pi^{6}-90\pi^{2})k +
576(\pi^{6}-90\pi^{2})k^{2}\over 2430 \zeta(5)} y^{-2}+ {\cal
O}(y^{-3})\Big).}}  The additional contributions to the partition
function in this expression have two notable features. First, and
most importantly, they are not zero. Thus, as described above, the
partition function truly cannot be represented as a sum of
exponentials. Second, they differ qualitatively from the leading
$y^{0}$ term: the additional coefficients appearing here are
positive and irrational, rather than negative and integer.

\subsec{$\hat n\ne 0$ Modes}

Now we will  consider the $\hat n\ne 0$ terms. For $\mu=0$ and
$\hat n \ne 0$, the integral \idef\ is a $K$-Bessel function
\eqn\inhat{ I_{m,{\hat n}}(s,\kappa,0) = {2^{s+1}\pi^{2s+m}|{\hat
n}|^{s+m-1/2} \over m!\Gamma(s+m) }y^{s+m-1/2}K_{s+m-1/2}(2\pi
|{\hat n}|y).} The Kloosterman sum is now the general case of
Ramanujan's sum \eqn\asd{ S({\hat n},0;c) = \sum_{d\in
(\Z/c\Z)^*}e^{2\pi i {\hat n} d/c } = \sum_{\delta|{\hat n}}
\mu(\delta).} We will simply quote the answer for the sum over $c$
(see Sec. 2.5 of \Iwaniec)  \eqn\snhat{
\sum_{c=1}^{\infty}c^{-2(s+m)} S({\hat n},0;c) = {1\over \zeta
(2(s+m))} \sum_{\delta | {\hat n}}\delta^{1-2(s+m)}. }

Taking $s=1/2$ gives a Fourier coefficient of the regularized
partition function. Consider first the $m=0$ term.  For $\hat
n=0$, this was the dangerous term in the analytic continuation,
but for $\hat n\not=0$, it simply vanishes, because the
$\zeta(2s)$ in \snhat\ has a pole at $c=1/2$, causing the
Kloosterman sum to vanish, and -- unlike the $\hat n = 0$ case --
the integral \inhat\ is finite at $s=1/2$. The other terms are
non-zero, and give \eqn\enhat{ E_{\hat n}(1/2,k,0) =
\sum_{m=1}^{\infty} {2^{3/2}\pi^{m+1}|{\hat n}|^{m}\over m!
\Gamma(m+1/2)\zeta(2m+1)}
 \left(\sum_{\delta|{\hat n}}\delta^{-2m}\right) \sqrt{y} K_{m} (2\pi |{\hat n}| y).
} Each term in the sum over $m$ vanishes for large $y$ as
$e^{-2\pi |\hat n| y}$.

Unfortunately, when $\mu\ne0$ the coefficients in the expansion of the
partition function are more difficult to compute.  The problem is not the
integral \idef; these integrals are finite and can be evaluated
analytically, although we will not write the answer here.  The answer is a finite sum of Bessel
functions of the form appearing in \inhat, each of which is multiplied by a
polynomial in $y$.  As $y\to\infty$, these Bessel functions cause the
integrals vanish as $e^{-2\pi |\hat n| y}$, just as in the
$\mu=0$ case described above.

However, when $\mu\ne0$, the sum over $c$ \eqn\csum{
\sum_{c=1}^{\infty} c^{-2(m+s)} S({\hat n},\mu,c), } though of
considerable number-theoretic interest, cannot be expressed in
terms of familiar number-theoretic functions such as the Riemann
zeta function.  Analytic properties of these sums have been
extensively studied. We will simply quote the relevant results. It
has been shown that the sum \csum\ defines a meromorphic function
on the complex $s$ plane. This function is essentially the Selberg
zeta function associated to the modular domain $D = {\eusm
H}/SL(2, \Z)$.  When we take $s=1/2$, the function \csum\ remains
regular for elementary reasons if $m>0$. Indeed, one can see
directly that the sum converges for these values, by noting from
the definition of the Kloosterman sum \klooster\ that $|S({\hat
n},\mu,c)|\le c$. To understand the case $m=0$, one needs deeper
results that can be found in Chapter 9 of \Iwaniec\ (and were
described to us by P. Sarnak). The key result is that the only
poles of this sum, or of its analogs for other congruence
subgroups of $SL(2,\Z)$, are at the points $s+m=1/2 + i t_{j}$,
where $t^{2}_{j}-1/4$ is one of the discrete eigenvalues of the
hyperbolic Laplacian $\Delta=-y^2(\partial_x^2+\partial_y^2)$ on
$D$.
 A pole at $s=1/2$, $m=0$, will therefore arise precisely if
 $\Delta$
has a discrete eigenvalue at $\lambda=1/4$. It is a happy fact
that no such eigenvalue exists  -- the smallest discrete
eigenvalue of the Laplacian on $D$ is of order $\lambda_{1}\approx
90$. Therefore, the sum \csum\ may be analytically continued to
give a finite value at $s=1/2$ for all values of $m$, including
the dangerous case $m=0$. In addition, simple bounds suffice to
show that there is no problem with the sum over $m$.

This argument shows that all $\hat n\ne 0$ Fourier coefficients of
the partition function \zsum\ are finite.  Together with the
results of the previous subsection for $\hat n=0$, this allows us
to conclude that the regularization scheme described in Sec. 3.1
provides a finite answer for the partition function \zsum. We
conjecture  that, just as for $\hat n=0$, the results are not
compatible with a Hilbert space interpretation.

\subsec{Aside: the Tree Level Sum over Geometries}

For comparison, we will now consider the sum which arises if we
neglect the one-loop contribution to the path integral described
in Sec. 2.2.  As we will see, in this case the answer is if
anything even worse.

The sum over geometries, if we take account of only the classical
action and not the one-loop correction, is \eqn\asd{Z^*(\tau) =
\sum_{c,d}\exp\{2\pi k\, {\rm Im}\, \gamma \tau\}.} This sum is
quadratically divergent, since at large $c$ and $d$ the summand
approaches one.  Before attempting to regularize the sum, let us
compare it to the ``correct'' sum \zzsum\ that does include the
one-loop correction.  The ``correct'' formula has a factor of
$1/\sqrt{{\rm Im}\,\tau}|\eta|^2$ that is outside the summation.
This factor is certainly important, but it does not affect whether
the sum converges.  There is also a factor of $|1-q|^2$ inside the
sum.  This factor also turned out in the above analysis to be less
important than it may have appeared; we simply expanded it as
$1-q-\bar q+q\bar q$, and wrote the partition function as a sum of
four terms.  All four terms were similar, and nothing particularly
nice happened in adding them up.

The factor in \zzsum\ that actually is important, resulted from
the one-loop correction, and has no analog in the ``naive'' sum
\asd\ is the innocent-looking factor of $\sqrt{{\rm
Im}\,\tau}|_\gamma$ that is {\it inside} the summation.  Because
of this factor, we had to evaluate our Poincar\'e series at
$s=1/2$.

This factor is absent in eqn. \asd, so now, if we try to define
the naive sum  $Z^*$ by introducing a parameter $s$ as before, we
will have to evaluate the resulting function at $s=0$. In fact,
the necessary $s$-dependent function was already introduced in
eqn. \poincare; $Z^*(\tau)$ is formally given by the function
$E(s,k,0)$ at $s=0$: \eqn\asd{Z^*(\tau) = \lim_{s\to 0}E(s,k,0).}
However, as an analytic function in $s$, $E(s,k,0)$ has a pole at
$s=0$. To see this, consider the expansion \poincaree\ of the part
of $E(s,k,0)$ which is constant in $x$. The $m=0$ term in this sum
vanishes, because of the pole in $\Gamma(s)$ at $s=0$.  The $m=1$
term gives \eqn\asd{E(s,k,0) = \sqrt{\pi}{\zeta(1+2s)\Gamma(s+1/2)
\over \zeta(2+2s) \Gamma(1+s)} + {\cal O}(y^{-1})} which has a
pole at $s=0$ coming from the harmonic series $\zeta(1)=\infty$.

Hence, without the one-loop correction, the divergence of the sum
over geometries becomes more serious.  Of course, even if we could
make sense of the function $Z^*$, we still might have trouble
giving it a Hilbert space interpretation.

One might wonder what happens if we modify the definition of $Z^*$
to include a finite subset of the Brown-Henneaux states -- for
example, the states of negative energy.  This means that before
summing over geometries, we multiply the exponential of the
classical action by a polynomial  $\sum_{n,m=0}^t a_{n,m}q^n\bar
q^m$. The sum over geometries is a sum of terms each of which is
similar to what was just described, with one such term for each
monomial $q^n\bar q^m$.  Each individual monomial will contribute
a pole at $s=0$, and generically this will survive in the sum.

\newsec{Possible Interpretations}

So far, we have analyzed the sum of known contributions to the
partition function of pure three-dimensional gravity.  As we
learned in Sec. 3, the resulting function cannot be interpreted as
$\Tr\,\exp(-\beta H)$ for any Hilbert space operator $H$.

We will now address the question of how to interpret this result.
The most straightforward interpretation is to take the result at
face value.  Three-dimensional pure gravity may not exist as a
quantum theory; to get a consistent theory, it may be necessary to
complete it by adding additional degrees of freedom, and there may
be no canonical way to do this.

The other possibility is that some unknown contributions to the
partition function should be added to the terms that we have
evaluated.  Here all sorts of speculations are possible. We will
consider two quite different possibilities.

\subsec{Cosmic Strings}

\lref\mms{J. Maldacena, J. Michelson, and A. Strominger, ``Anti de
Sitter Fragmentation,'' JHEP 9902:011 (1999), hep-th/9812073. }

\lref\seiwit{N. Seiberg and E. Witten,  ``The D1/D5 System and
Singular CFT,'' JHEP 9904:017 (1999), hep-th/9903224.}

\lref\krasnov{K. Krasnov, ``On Holomorphic Factorization In
Asymptotically AdS 3D Gravity,'' Class. Quant.Grav. {\bf 20}
(2003( 4015-4042,  hep-th/0109198.}

\lref\whox{G. H\"ohn, ``Selbstduale Vertexoperatorsuperalgebren
und das Babymonster,''
 Ph.D. thesis  (Bonn 1995),
 Bonner Mathematische Schriften 286 (1996), 1-85, English translation, arXiv:0706.0236.}

 \lref\gaiottoxi{ D. Gaiotto and X. Yin, ``Genus Two Partition Functions of Extremal Conformal Field
 Theories,''
 JHEP 0708:029 (2007), arXiv:0707.3437.}

 \lref\gaberdiel{M. Gaberdiel, `` Constraints on Extremal Self-Dual
 CFTs,'' arXiv:0707.4073.}

\lref\xi{X. Yin,  ``Partition Functions of Three-Dimensional Pure
Gravity,'' arXiv:0710.2129.}

 \lref\gaiotto{D. Gaiotto, to appear.}

\lref\flm{I. B. Frenkel, J. Lepowsky, and A. Meurman, {\it Vertex
Operator Algebras And The Monster} (Academic Press, Boston,
1988).}

\lref\knopp{M. I. Knopp, ``Rademacher on $J(\tau)$, Poincar\'e
Series Of Nonpositive Weights and the Eichler Cohomology,''
Notices Am. Math. Soc. {\bf 37} (1990) 385-93.}

\lref\knopptwo{M. I. Knopp, ``On The Fourier Coefficients Of Cusp
Forms Having Small Positive Weight,'' Proc. Symp. Pure Math. {\bf
49} (1989) part 2, 111-127.}

 Known consistent models of $2+1$-dimensional gravity with negative
cosmological constant arise from string theory.  For example, a
famous class of models comes from Type IIB superstring theory on
$\AdS_3\times S^3\times X$, where $X $ is either a K3 surface or a
four-torus.

In these models, the dimensionless ratio $k=\ell/16 G$ is never a
variable parameter, but always takes quantized values determined
by fluxes that are chosen in the compactification.  (The fact that
$\ell/G$ is not continuously variable is actually  \WittenKT\ a
more general consequence of the Zamolodchikov $c$-theorem applied
to the boundary CFT.)  Moreover, it is always possible to have
domain walls across which the fluxes jump.  The domain walls are
constructed from suitably wrapped branes.

In $2+1$ dimensions, a domain wall has a $1+1$-dimensional
world-volume and so can be viewed as a cosmic string.  The
existence of these cosmic strings makes the models much more
unified, as regions with different fluxes can appear as different
domains in a single spacetime.

The usual $\AdS_3\times S^3\times X$ models  have supersymmetric
moduli, and the values of the string tension depend on these moduli.
There is a particularly interesting supersymmetric value of the
string tension at which ``long strings'' become possible.    These
are strings that can expand to an arbitrarily large size at only a
finite cost of energy \refs{\mms,\seiwit}. When long strings exist,
the energy spectrum is continuous above a certain minimum energy,
and the partition function $\Tr\,\exp(-\beta H)$ therefore diverges
for all $\beta$.

The numerical value of the long string tension and of the
excitation energy above the ground state beyond which the spectrum
is continuous are proportional to the jump in $k=\ell/16 G$ in
crossing the string. The threshold excitation energy is of order 1
(above the ground state at energy $-2k$) if the jump in $k$ is of
order 1.

Since  well-established models of three-dimensional gravity have
such cosmic strings, perhaps they also present in minimal
three-dimensional gravity, if it exists. In anything that one would
want to call pure gravity, the string tension $T$ measured in units
of $1/\ell^2$ must go to infinity as $k\to\infty$. Otherwise, the
cosmic strings would contribute excitations at the $\AdS_3$ scale,
and one would describe the model as a theory of three-dimensional
gravity plus matter.  We have no idea if the string tension should
be proportional to $k$ (as one might expect for solitons), to
$k^{1/2}$ (as for $D$-branes), etc.  If the jumps in $k$ are
correctly matched with $T$, then the strings are long strings and
the partition function $\Tr\,\exp(-\beta H)$ that we have been
trying to compute in this paper is actually divergent. If the jumps
in $k$ are smaller than this, then the partition function converges,
but to compute it might involve corrections from the strings.

If one thinks that the strings should be long strings, then the
requirement $T>>\ell^2$ means that the jumps in $k$ in crossing
strings are much greater than 1.  There is some tension between this
and the proposal in \WittenKT\ that $k$ can take any integer value.
If all allowed values of $k$ are connected by strings or domain
walls (as in the known string/$M$-theory models), then the fact that
the jumps in $k$ are large means that the allowed values of $k$ are
sparse.

\subsec{Doubled Sum Over Geometries}

The scenario just described is obviously rather speculative, but
at least it has the virtue of underscoring our point that the
range of conceivable unknown contributions to the $\AdS_3$
partition function is quite large.

We will now describe a quite different scenario.  To motivate it,
we return to the formula \gopher\ for the classical action of the
basic spacetime $M_{0,1}$:
 \eqn\yngo{I = -4\pi k {\rm
Im}\,\tau.}  We write more explicitly \eqn\tyngo{I=2 \pi i
k(\tau-\bar\tau).} Hence the classical approximation  for the
contribution of this spacetime to the partition function is
$\exp(-I)=\exp(-2\pi i k (\tau-\bar \tau))=q^{-k}\bar q^{-k}$.  We
notice that this is locally the product of a holomorphic function
of $k$ and an antiholomorphic function, and is globally such a
product if $k$ is an integer (ensuring that $q^{-k}$ is
single-valued). The one-loop correction preserves this factorized
form, and therefore the formula \felox\ for the exact partition
function $Z_{0,1}$ asociated with $M_{0,1}$ has the same
properties; in fact, $Z_{0,1}=F_k(q)F_k(\bar q)$, with
\eqn\melf{F_k(q)=q^{-k}\prod_{n=2}^\infty (1-q^n)^{-1}.}

To the extent that known formulations of three-dimensional gravity
are valid, this sort of factorization holds for the contribution
to the partition function  of any classical geometry. See
\krasnov\ for a detailed example. The gauge theory description
\refs{\townsend,\witten} of three-dimensional gravity gives a
natural explanation of this. With negative cosmological constant,
in Lorentz signature, the gauge group is $SL(2,\R)\times
SL(2,\R)$; the theory is a product of two decoupled $SL(2,\R)$
theories, associated respectively with left- and right-moving
modes in the boundary CFT, and this corresponds to holomorphic
factorization in the Euclidean form of the theory.

In \WittenKT, it was suggested that the {\it exact} partition
function of pure three-dimensional gravity is holomorphically
factorized. It was observed that if this is the case, and the
Brown-Henneaux spectrum is exact until one gets above the
classical black hole threshold, then the partition function can be
determined uniquely. Dual CFT's consistent with the necessary
spectrum have been called extremal CFT's \whox.  In subsequent
work, it has been shown that the genus 2 partition function of
such a CFT can be uniquely and consistently determined \gaiottoxi,
and there has been some work comparing it to what would be
expected from three dimensions \xi, but on the other hand an
interesting but slightly technical argument has been given which
may show that extremal CFT's do not exist \gaberdiel.  Also, it
has been argued \gaiotto\ that extremal CFT's, if they exist,
generally do not have monster symmetry for $k>1$, in contrast to
what happens \flm\ for $k=1$.

Now let us discuss holomorphic factorization in view of the sum
over geometries.  Associated to an element
\eqn\londly{\gamma=\left(\matrix{ a & b\cr c & d\cr}\right)} of
$SL(2,\Z)$ is a classical spacetime $M_{c,d}$.  Its action is
obtained by applying a modular transformation to \tyngo:
\eqn\ondly{I_{\gamma}(\tau)=2\pi i
k\left(\gamma\tau-\gamma\bar\tau\right).}
 As usual, $\gamma\tau=(a\tau+b)/(c\tau+d)$,
$\gamma\bar\tau=(a\bar\tau+b)/(c\bar\tau+d)$.
 The
partition function of the manifold $M_{c,d}$ is
\eqn\geug{Z_{c,d}=\left.F_k(q)\right|_\gamma \left. F_k(\bar
q)\right|_\gamma,} and is holomorphically factorized just like
$Z_{0,1}$.

\font\teneurm=eurm10 \font\seveneurm=eurm7 \font\fiveeurm=eurm5
\newfam\eurmfam
\textfont\eurmfam=\teneurm \scriptfont\eurmfam=\seveneurm
\scriptscriptfont\eurmfam=\fiveeurm
\def\eurm#1{{\fam\eurmfam\relax#1}}
\def\W{{\eurm W}}
However, when we sum over geometries to evaluate the partition
function \eqn\bondly{Z=\sum_{\gamma\in \W}F_k(q)|_\gamma F_k(\bar
q)|_\gamma,} holomorphic factorization is lost. (Here $\W$ is set of
classical geometries $M_{c,d}$, isomorphic to the coset space
$PSL(2,\Z)/\Z$, with $\Z$ being the upper triangular subgroup of
$SL(2,\Z)$.) In fact, this formula is not simply a product of
holomorphic and antiholomorphic functions, but a sum of such
products, somewhat like the partition function of a rational
conformal field theory.  (However, in that case, each term appearing
in the sum is separately invariant under $T:\tau\to\tau+1$.)

\def\hat{\widehat}
Holomorphic factorization has been lost because the sum over
topologies is a common sum for left- and right-movers. What could
be added to restore holomorphic factorization?  This question has
a simple answer, though  whether the answer is really relevant to
three-dimensional gravity remains to be seen.

If we formally introduce separate topological sums for holomorphic
and antiholomorphic variables, defining an extended partition
function \eqn\expar{\hat Z=\sum_{\gamma,\gamma'\in
\W}\left.F_k(q)\right|_\gamma \left.F_k(\bar q)\right|_{\gamma'},}
then holomorphic factorization is restored (if the sum converges
or can be regularized in a satisfactory way).  After all, \expar\
can be written in the manifestly holomorphically factorized form
\eqn\xpar{\hat Z=\left(\sum_{\gamma\in \W}\bigl.
F_k(q)\bigr|_\gamma \right) \left(\sum_{\gamma'\in \W}\bigl.
F_k(\bar q)\bigr|_{ \gamma' }\right).} The question (apart from
regularizing these sums) is to interpret the double sum over
topologies.

What is the classical action corresponding to a given term in the
sum in \xpar?  We can answer this question by simply applying
separate modular transformations to holomorphic and
antiholomorphic variables on the right hand side of eqn. \tyngo.
The ``action'' of a ``classical solution'' that would lead to the
term in \expar\ labeled by a pair $\gamma,\gamma'$ must be
\eqn\zpar{I_{\gamma,\gamma'}=2\pi
i\left(\gamma\tau-\gamma'\bar\tau\right).} For
$\gamma\not=\gamma'$, this formula is not real, so it is not the
action of a real solution of the Einstein equations.

The most obvious way to try to interpret the formula is to
interpret it as the action of a complex-valued solution of the
Einstein equations.  As we noted in the introduction, naively
speaking the Euclidean path integral is a sum over three-manifolds
$M$ that obey the boundary conditions.  But there is no
established way to evaluate the contribution to the Euclidean
functional integral of a given $M$ unless $M$ admits a classical
solution of the equations of motion that one can expand around. It
is because of this that our first step in Sec. 2.1 was to classify
the classical solutions; what we classified were the solutions of
the usual real Einstein equations with Euclidean signature.

It seems strange to simply ignore three-manifolds that do not
admit classical solutions.  Are their contributions meaningless or
zero for some reason?

 One obvious
way to slightly generalize the usual framework is to consider the
complexified equations of motion -- the Einstein equations $R_{\mu
\nu}=-\Lambda g_{\mu\nu}$, defined in the usual way except with a
complex-valued but nondegenerate metric tensor $g_{\mu\nu}$. If
such a complex-valued solution is given, one can hope that
perturbation theory around it would make sense. In the present
theory, in which perturbation theory about a classical solution is
one-loop exact, one could hope to make sense of the contribution
of a complex saddle point (that is, a complex-valued solution of
the equations of motion) to the partition function.

Unfortunately, we have been unable to find a convincing family of
solutions of the complexified Einstein equations depending on the
pair $\gamma,\gamma'\in\W$.  Since this is the case, we think
another possibility is that the nonperturbative framework of
quantum gravity really involves  a sum not over ordinary
geometries in the usual sense, but over some more abstract
structures that can be defined independently for holomorphic and
antiholomorphic variables. (A similar idea is expressed in Sec.
3.1 of \xi.)  Only when the two structures coincide can the result
be interpreted in terms of a classical geometry.

\bigskip\noindent{\it ``Seeing'' The Non-Classical Geometries}

None of this is terribly convincing -- no more so, certainly, than
the discussion of cosmic strings in Sec. 4.1.  However, we can
describe one fact that offers some slight encouragement.

Let us examine the classical limit of the partition function, by
which we mean the limit of $k\to\infty$ with fixed $\tau$. First
we will consider the ordinary partition function $Z$, defined by
summing over ordinary geometries, and then we will consider the
extended partition function $\hat Z$, defined in \expar\ by a
double sum whose meaning is unclear.

In the classical limit, the sum over geometries is dominated by
the geometry that has the most negative classical action.  In
other words, we must pick $\gamma$ to minimize the quantity
\eqn\yolpo{I_\gamma=2\pi i( \gamma\tau -\gamma\bar\tau).} We can
also write \eqn\polpo{I_\gamma={\rm Re}\,\left(4\pi
i\gamma\tau\right)={\rm Re}\,\left(-4\pi i \gamma\bar\tau\right).}
For generic $\tau$, the maximum occurs for a unique $\gamma$.

There actually is a complete democracy between the different choices
of $\gamma$, because they are all obtained from each other by
$SL(2,\Z)$ transformations.  Each $\gamma$ dominates the partition
function (in the large $k$ limit) for $\tau$ in a suitable region of
the upper half plane.  The thermal $\AdS$ manifold $M_{0,1}$
dominates in the usual fundamental domain $|\tau|>1$, $|{\rm
Re}\,\tau|<1/2$.  Across the arc $|\tau|=1$, ${\rm Re}\,\tau\leq
1/2$, there is a Hawking-Page phase transition to a region dominated
by $M_{1,0}$, which is the Euclidean black hole. The upper half
plane is tesselated in phases dominated by different classical
solutions. Details are further discussed in Sec. 6.

Now let us consider the extended partition function $\hat Z$.  The
action, given in eqn. \zpar\ now depends on the pair
$\gamma,\gamma'$, and is complex.  What we want to minimize is the
real part of the action: \eqn\bpar{{\rm
Re}\,I_{\gamma,\gamma'}={\rm Re}\,\left(2\pi
i(\gamma\tau-\gamma'\bar\tau)\right)={\rm Re}\,\left(2\pi
i\gamma\tau\right)+{\rm Re}\,\left(-2\pi
i\gamma'\bar\tau\right)={\rm Re}\,\left(2\pi
i\gamma\tau\right)+{\rm Re}\,\left(2\pi i\gamma'\tau\right).} The
minimum is always at $\gamma=\gamma'$.  In fact, we can use eqn.
\polpo\ to show that \eqn\cpar{{\rm
Re}\,I_{\gamma,\gamma'}={1\over
2}\left(I_\gamma(\tau)+I_{\gamma'}(\tau)\right).} Whatever is the
minimum with respect to $x$ of $I_x(\tau)$, the minimum of ${\rm
Re}\, I_{\gamma,\gamma'}(\tau)$ is at $\gamma=\gamma'=x$.

The conclusion is that, even if non-classical geometries exist and
the correct partition function is a sum over non-classical
geometries as well as classical ones, the semiclassical limit of
large $k$ is always dominated by a classical geometry.
Non-classical geometries exist, but they never dominate the
semiclassical limit.

What do we have to do to ``see'' a non-classical geometry in the
semiclassical limit?  Naturally, we have to ask a non-classical
question.  The partition function as we have defined it so far is
$Z(\tau)=\Tr\,\exp(2\pi i\tau L_0-2\pi i \bar\tau\tilde L_0)$,
where $\tau$ is a point in the upper half plane $\eusm H$ and
$\bar\tau$ is its complex conjugate. In terms of
$\tilde\tau=-\bar\tau$, which also takes values in $\eusm H$, we
have $Z(\tau)=\Tr\,\exp(2\pi i\tau L_0+2\pi i\tilde\tau\tilde
L_0)$. Now instead of defining $\tilde\tau$ as $-\bar \tau$, let
us relax this condition and simply think of $\tilde\tau$ as a
second point in $\eusm H$.\foot{The partition function of a theory
in which the high energy density of states is given by the entropy
of a BTZ black hole remains convergent after this analytic
continuation to general $\tau,\tilde\tau\in \eusm H.$} Then we
should rewrite \cpar\ in the form \eqn\icpar{{\rm
Re}\,I_{\gamma,\gamma'}={1\over
2}\left(I_\gamma(\tau)+I_{\gamma'}(\tilde\tau)\right).} Now it is
clear that given any pair $\gamma,\gamma'$, we can pick the pair
$\tau,\tilde\tau$ so that the extended partition function $\hat Z$
is dominated in the semiclassical limit by $\gamma,\gamma'$. We
simply pick $\tau$ so that $I_x(\tau)$ is minimized for
$x=\gamma$, and $\tilde\tau$ so that $I_x(\tilde\tau)$ is
minimized for $x=\gamma'$.

Thus, for each non-classical geometry defined by a pair $\gamma,
\gamma'$, we can find a question that that geometry dominates in
the semiclassical limit.  But to do this, we have to ask a rather
exotic question that itself depends on a rather unusual analytic
continuation.

\newsec{Black Hole Entropy and its Corrections}

In this and the following section we will discuss a few
implications of the assumption of holomorphic factorization.  In
this section we will discuss black hole entropy, and how it can be reproduced in holomorphically factorized theories.

In Sec. 2, we computed the perturbative corrections to the saddle
point action for the geometry $M_{0,1}$.  Using a modular
transformation, this leads to a new formula for the subleading
corrections to the entropy of the BTZ black hole.  We will compare
these new subleading corrections with those that occur in the
extremal holomorphic partition functions of \WittenKT.

\subsec{Subleading Corrections to the Bekenstein-Hawking Formula}

As we noted in Sec. 2, the geometry we called $M_{1,0}$ is just the
Euclidean continuation of the BTZ black hole.  The modular parameter
$\tau$ is related to the Hawking temperature $\beta^{-1}$ and
angular potential $\theta$ of the black hole, by
$\tau=\theta+i\beta$. The black hole partition function is found by
applying the modular transformation $\tau\to-{1/\tau}$ to the
expression \zelox\ for the partition function $Z_{0,1}$ of thermal
AdS.  This gives the one-loop corrected partition function for the
Euclidean BTZ black hole \eqn\bhpartt{ Z_{1,0} = {\cal Z}(\tau)
\bar{\cal Z}({\bar \tau})} where we have defined the holomorphic
piece of the BTZ partition function \eqn\bhpart{{\cal
Z}(\tau)={q_{-}^{-(k-1/24)}(1-q_{-})\over \eta(-1/\tau)}.}
In this formula we have defined ${q_{-}} = e^{-2\pi i/\tau}$. It
will also be useful to write ${\cal Z}(\tau)$ as  \eqn\zholexp{{\cal
Z}(\tau) = \sum_{\Delta=-k}^{\infty} C_{\Delta} q_{-}^{\Delta}} where
the coefficients $C_{\Delta}$ are \eqn\fare{C_{\Delta} = p(\Delta' +
k) - p(\Delta' + k-1).}  Here $p(N)$ denotes the number of
partitions of the integer $N$.

Equation \bhpart\ encodes various quantum corrections
to the thermodynamic properties of the BTZ black hole. Equation \bhpart\ is a canonical ensemble partition function, so one may, for
example, compute the black hole entropy using the usual
formula\eqn\sis{\eqalign{ S(\beta,\theta) &= {\rm log}~Z_{1,0} - \beta^{}~Z_{1,0}^{-1}~ {\partial Z_{1,0}\over \partial \beta}}}
where $\tau=\theta+i\beta$.
This is a canonical ensemble entropy, evaluated at fixed temperature $\beta^{-1}$ and angular potential $\theta$.

For the purpose of comparing with CFT predictions, however, it is more useful to compute the microcanonical entropy, which counts the number of states $N(M,J)$ at fixed energy $M$ and angular momentum $J$.  The energy $M$ and angular momentum $J$ of a state are related to left and right-moving dimensions of the state by\foot{We are working in AdS units, with $\ell=1$.}
\eqn\asd{\eqalign{
M &= \Delta + \bar \Delta \cr J &= \Delta-\bar \Delta
.}}So we may write this density of states as $N(\Delta,\bar \Delta)$.  This density of states is computed from the partition function $Z_{1,0}(\tau, \bar \tau) $ by a pair of Laplace transforms
\eqn\nis{ N(\Delta,\bar\Delta) = 
\left( \int_{i\epsilon-\infty}^{i\epsilon+\infty} d\tau\right)
\left( \int_{i\epsilon-\infty}^{i\epsilon+\infty} d\bar\tau \right)
q^{-\Delta}{\bar q}^{-\bar \Delta} Z_{1,0}(\tau,\bar \tau) .}
These are the usual Laplace transforms that appear when going from canonical to microcanonical ensemble.  We should emphasize that in this expression
$\tau$ and $\bar \tau$ should be regarded as
independent variables.

We should note that, since \bhpart\ is a
semiclassical partition function around a black hole background, it
will not have an exact interpretation as a quantum mechanical trace.  Moreover, the quantization of $\Delta$ and $\bar \Delta$ will not be
visible in this approximation.\foot{To be more precise, in an exact
theory with finite entropy, $N(\Delta,\bar\Delta)$ will be a sum of
delta functions of $\Delta,\bar\Delta$ multiplying positive
integers; but the semiclassical approximation does not have this
form.} So equation \nis\ should be thought of only as a
semi-classical approximation to the number of states with dimension
$(\Delta,\bar \Delta)$ in the exact quantum theory. Still, we want
to explore this semiclassical formula.

The microcanonical entropy $S(\Delta,\bar \Delta)$ is just
\eqn\asd{S(\Delta,\bar\Delta) = {\rm log}~{N}(\Delta, \bar \Delta).}Of course, since $Z_{1,0}$ is holomorphicaly factorized,
we may write \eqn\asd{N(\Delta,\bar\Delta) = {\cal N}(\Delta) {\cal N}(\bar \Delta),~~~~~~S(\Delta,\bar \Delta)
= {\cal S}(\Delta) + {\cal S}(\bar \Delta)} where \eqn\asd{
 {\cal N}(\Delta) = \int_{i\epsilon-\infty}^{i\epsilon+\infty}d\tau \,q^{-\Delta}
 {\cal Z}(\tau) .
 }
 We may use the expansion \zholexp\ to get
\eqn\nderiv{\eqalign{
{\cal N}(\Delta) &=  \sum_{\Delta'=-k}^{\infty} C_{\Delta}
\int^{i\epsilon+\infty}_{i\epsilon-\infty} d\tau q^{-\Delta}q_{-}^{\Delta'} \cr
&=  \sum_{\Delta'=-k}^{\infty}C_{\Delta}\int^{i\epsilon+\infty}_{i\epsilon-\infty} d\tau \exp\left\{-2\pi i \left(\Delta \tau + {\Delta'\over \tau}\right) \right\}
.}}
To do this integral, recall the following representation of the Bessel function
as a contour integral: \eqn\asd{I_1(z) =
{1\over 2\pi i}\oint t^{-2} e^{(z/2)(t + t^{-1}) } dt }
where the contour encloses the origin in a counterclockwise direction. 
Taking $t\to1/t$ and letting $z= 4\pi \sqrt{-\Delta \Delta'}$, this becomes the integral appearing in \nderiv.  We end up with the following formula for the microcanonical entropy:
\eqn\nis{
{\cal N}(\Delta) = e^{{\cal S}(\Delta)}={2\pi} \sum _{\Delta'=-k}^{\infty}C_{\Delta} \sqrt{-\Delta'\over \Delta} I_{1}
(4\pi \sqrt{-\Delta \Delta'}) }where $C_{\Delta}$ was defined in \fare.
Formula \nis\ is the main result of this section.

In the semiclassical approximation, this formula will be dominated by the term with
$\Delta'=-k$.  We may then use the asymptotic formula for the Bessel function\eqn\bessel{I_1(z) = {1\over \sqrt{2 \pi
z}}e^z\left(1-{3\over 8} z^{-1} + ... \right),~~~~~{\rm at}\
z\to\infty}
to get\eqn\asd{{\cal S}(\Delta) = {\rm log}~{\cal N}(\Delta) =
4\pi\sqrt{k\Delta} +{1\over 4} {\rm log}~k -{3\over 4} {\rm log}~\Delta - {1\over 2}
{\rm log}~2  +\dots}The first term is the usual Bekenstein-Hawking term, proportional to the area of the BTZ black hole.  The other
terms are logarithmic corrections that typically appear when the entropy is computed in microcanonical,
as opposed to a canonical, ensemble.  They were computed for the BTZ black hole in \CarlipNV.   

Of course, equation \nis\ contains many interesting subleading
terms in addition to these logarithmic terms.  Let us consider the terms in \nis\ with $\Delta>-k$.  For large values of $\Delta/k$, which is to say for black holes whose radius is large in AdS units,
these terms are exponentially subleading.  To see this, let us compare the $\Delta=-k$ and $\Delta=-k+1$
terms in the sum using the asymptotic behavior of the Bessel function \bessel.  We find \eqn\asd{{I_1(4\pi \sqrt{k \Delta} ) \over I_1(4\pi
\sqrt{(k-1)\Delta})} \sim e^{2\pi \sqrt{\Delta/k}},~~~~~{\rm as}\ \Delta
k\to\infty .}Thus for $\Delta/k$ large the sum is dominated by the
$\Delta=-k$ term plus terms which are exponentially small in $\sqrt{\Delta/k}$.  For black holes
whose size is of order the AdS scale, these additional terms can become relevant.

We should make one more important comment about the structure of the subleading terms in \nis.   The terms with $\Delta'<0$ are qualitatively of the same form as the leading contribution described above.  The terms with $\Delta'>0$ are qualitatively different: for these terms the factors of $\sqrt{-\Delta'}$ appearing in \nis\ are imaginary, and we must worry, for example, about which branch of the solution \nis\ we should take.  At this point one might wonder about the physical interpretation and implications of these funny terms. However, as we will see below, it is precisely these unusual-looking terms which are absent in the correct microscopic computation of the black hole entropy.  So we will not worry too much about their appearance here.

Our goal is to understand to what extent the microcanonical entropy formula \nis\ can be reproduced in a
holomorphically factorized partition function.  Of course, we could attempt to reproduce the canonical
formula \sis\ instead, as these two expressions contain precisely the same information.   But we will find
the problem of computing \nis\ to be technically simpler.  Our conclusion will be that the terms in \nis\ with $-k \le \Delta' <0$ are reproduced in the full partition function.

Before proceeding, let us ask under what circumstances we expect the formula \nis\ for the entropy to be a
good approximation to the number of states of the exact quantum theory.  Of course, we must take $k$ to be
large, in order for the semi-classical approximation to be valid.  In addition, we must take $\Delta$ to be
large so that the corresponding black hole is large in Planck units and dominates the entropy of the system.

\subsec{The Rademacher Expansion for Holomorphic Partition Functions}

Let us now consider a modular-invariant, holomorphic CFT with
central charge $c=24k$.  For such CFTs, $k$ must be an integer, so
the holomorphic part of the partition function can be expanded as
\eqn\zexact{ Z(\tau) = \sum_{\Delta=-k}^\infty F_{\Delta}
q^\Delta}where the $F_{\Delta}$ are some positive integers. Modular
invariance fixes $Z(\tau)$ in terms of the finite set of
coefficients $\{ F_{\Delta} | -k \le \Delta \le 0 \}$. Once these
polar coefficients are fixed, an explicit formula can then be given
for $Z(\tau)$ as a polynomial in the $j$-invariant, although we will
not write this formula here. The important point is that all of the
coefficients $F_{\Delta}$ with $\Delta>0$ are fixed in terms of the
coefficients $F_{\Delta}$ with $\Delta\le0$.  The formula for the
$F_{\Delta}$ with $\Delta >0$ in terms of these polar terms is \eqn\rade{F_{\Delta} =
2\pi \sum_{-k\le\Delta'<0} \sqrt{|\Delta'| \over \Delta} F_{\Delta'}
\sum_{n=1}^\infty {1\over n} S(\Delta,|\Delta'|, n) I_1 \left({4\pi
\over n} \sqrt{\Delta |\Delta'|}\right) }where $S(\Delta,\Delta',n)$
is the Kloosterman sum defined in \klooster.  This is a convergent series expansion of $F_{\Delta}$.  Expansions of the form
\rade\ are  known as Rademacher expansions. Rademacher expansions
were first applied to three-dimensional gravity in \farey.

In \WittenKT, a conjecture was made regarding the form of the
partition function \zexact. The conjecture is that the theory
contains no primary fields with $-k<\Delta\le0$.  CFTs
with this property are known as extremal CFTs, since in a sense this
is the most extreme conjecture possible: modular invariance forces
one to include additional primaries with $\Delta>0$.
In terms of the coefficients $F_{\Delta}$, the conjecture of
\WittenKT\ is that \eqn\asd{ F_{\Delta} = C_{\Delta}~~~~~{\rm for}~
-k<\Delta \le 0.} Here $C_{\Delta}$ are the perturbative coefficients
defined in \fare.

The extremal partition function (or any similar function in which
the number of primary fields of dimension $\leq k$ is not too large)
is consistent with the leading Bekenstein-Hawking term in the black
hole entropy for $\Delta,k$ large. We are now in a position to ask
whether subleading terms given by \nis\ are reproduced as well.
First, we note that the sums \rade\ and \nis\ are quite similar. In
particular, when $n=1$ the Kloosterman sum is 1.  In this case the summand in \rade\ coincides precisely with that in \nis, at least when $\Delta'<0$. Thus the Rademacher expansion \rade\ contains the first $k$ terms in the one-loop corrected black hole entropy \nis.
This is the main result of this section.

There are two important differences between \rade\ and \nis.
The first is that the sum in \nis\ is over
$-k<\Delta'<\infty$, while the sum \rade\ is only over $-k<\Delta'<0$.  So the funny terms in \nis\ with $\Delta'>0$ are absent in the full microscopic entropy.  It would be
interesting to understand the physical interpretation of these terms, but we will not attempt to do so here.

The second important difference is that the Rademacher expansion \rade\ includes terms with $n>1$.  These terms
provide exponentially small corrections to the black hole entropy.
To see this, we may again use the asymptotic behavior of the Bessel function \bessel.  For black holes with large $\Delta$ the ratio of the
$n=1$ and $n=2$ terms in the sum \rade\ goes like \eqn\asd{{I_1(2\pi
\sqrt{\Delta |\Delta'|}) \over I_1(4\pi \sqrt{\Delta |\Delta'|})}
\sim e^{-2\pi \sqrt{\Delta |\Delta'|}}.} So at large $\Delta$, the
$n>1$ terms in the Rademacher expansion are exponentially small in
$\Delta$.  Again, it would be interesting to study the physical interpretation
of these terms.

\newsec{The Hawking-Page Phase Transition}

\lref\leeyang{C. N. Yang and T.-D. Lee, ``Statistical Theory Of
Equations Of State And Phase Transitions. I. Theory Of
Condensation,'' Phys. Rev. (1952) {\bf 87} 404-9.}
\lref\fisher{M.
Fisher, in {\it Lectures In Theoretical Physics} VIIC (University
of Colorado Press, Boulder, 1965).}

In this section we address an apparent puzzle posed by holomorphic
factorization and the Hawking-Page transition \hawkingpage. As we
will see, the resolution to this puzzle is that if holomorphic
factorization is valid, then  the Hawking-Page phase transition is
described by a condensation of Lee-Yang zeroes of the partition
function \refs{\leeyang,\fisher}.  We examined the zeroes of the
partition function as a result of a question from M. Kaneko, and
the zeroes we find are similar to what have been obtained in
investigations \refs{\Rankin, \AKN} of certain somewhat similar
modular functions.

The  partition function $Z(\tau)$ computes a canonical ensemble
partition function at fixed temperature ${\rm Im}~\tau$ and angular
potential ${\rm Re}~\tau$.  As described in Sec. 2, for each value
of $\tau$, an infinite family of classical geometries $M_{c,d}$ will
contribute to $Z(\tau)$.  The pair of integers $(c,d)$ labels an
element of the coset $SL(2,\Z)/\Z$, where $\Z$ denotes the shifts
$\tau\to\tau+n$. In Sec. 2.2, we computed the contribution of
$M_{c,d}$ to the partition sum $Z(\tau)$. If we choose an element
$\left(a~b\atop c~d\right)$ of $SL(2,\Z)$ corresponding to
$M_{c,d}$, the contribution to the partition function was given by
$Z_{0,1}({a\tau+b\over c\tau+d})$.

For a given value of $\tau$, one can ask which geometry $M_{c,d}$
has the largest contribution to $Z(\tau)$ in the semiclassical
limit $k\to\infty$. As explained in \refs{\MaldacenaBW,\farey} and
in Sec. 4.2, this question can be answered by maximizing the
classical action $-4\pi {\rm Im}\,(\gamma\tau)$ as a function of
$\gamma$.  If $\tau$ is in the usual fundamental domain
$|\tau|>1$, $|{\rm Re}\,\tau|\leq 1/2$, or any of its translates
by $\tau\to\tau+n$, then the answer is that the dominant classical
solution is $M_{0,1}$, which describes thermal AdS space.  For any
$\tau$, the dominant classical solution can be found by asking
 which value of $\gamma\in SL(2,\Z)$ will
take $\tau$ into the fundamental domain or one of its translates
under $\tau\to\tau+n$.  For example, $M_{1,0}$, whose Lorentzian
continuation describes a black hole in equilibrium with its Hawking
radiation,  dominates whenever there exists an $n$ such that
$-1/(\tau+n)$ lies in the fundamental domain. Roughly speaking, this
is a regime of low temperature and angular potential. More
complicated geometries $M_{c,d}$, which do not have simple
Lorentzian interpretations (as they do not admit an analytic
continuation to a real solution with Lorentz signature) dominate in
other regions of the upper half plane $\eusm H$. Putting this
together, we find that the phase diagram of three-dimensional
gravity, in the semiclassical limit, is given by a sort of
subtesselation of the usual tesselation of $\eusm H$ by fundamental
domains of $SL(2,\Z)$.

We get a subtesselation because certain boundaries between
fundamental domains can be crossed without any jump in the dominant
geometry (an example is that, for ${\rm Im}\,\tau>>1$, there is no
jump in crossing between domains related by $\tau\to\tau+n$).  The
usual tesselation of $\eusm H$ by fundamental domains is sketched in
Fig. 3a, and the subtesselation relevant to three-dimensional
gravity is sketched in Fig. 3b.

\doublefig{ a) The standard tesselation of the upper half
$\tau$-plane by $SL(2,{\Z})$ fundamental regions. b) The
subtesselation that represents the phase diagram of three
dimensional gravity. The phase boundaries, represented by solid
black arcs,  connect fixed points of $SL(2,\Z)$ of order 3. This
characterization was found by considering special cases and then
using modular invariance.}
{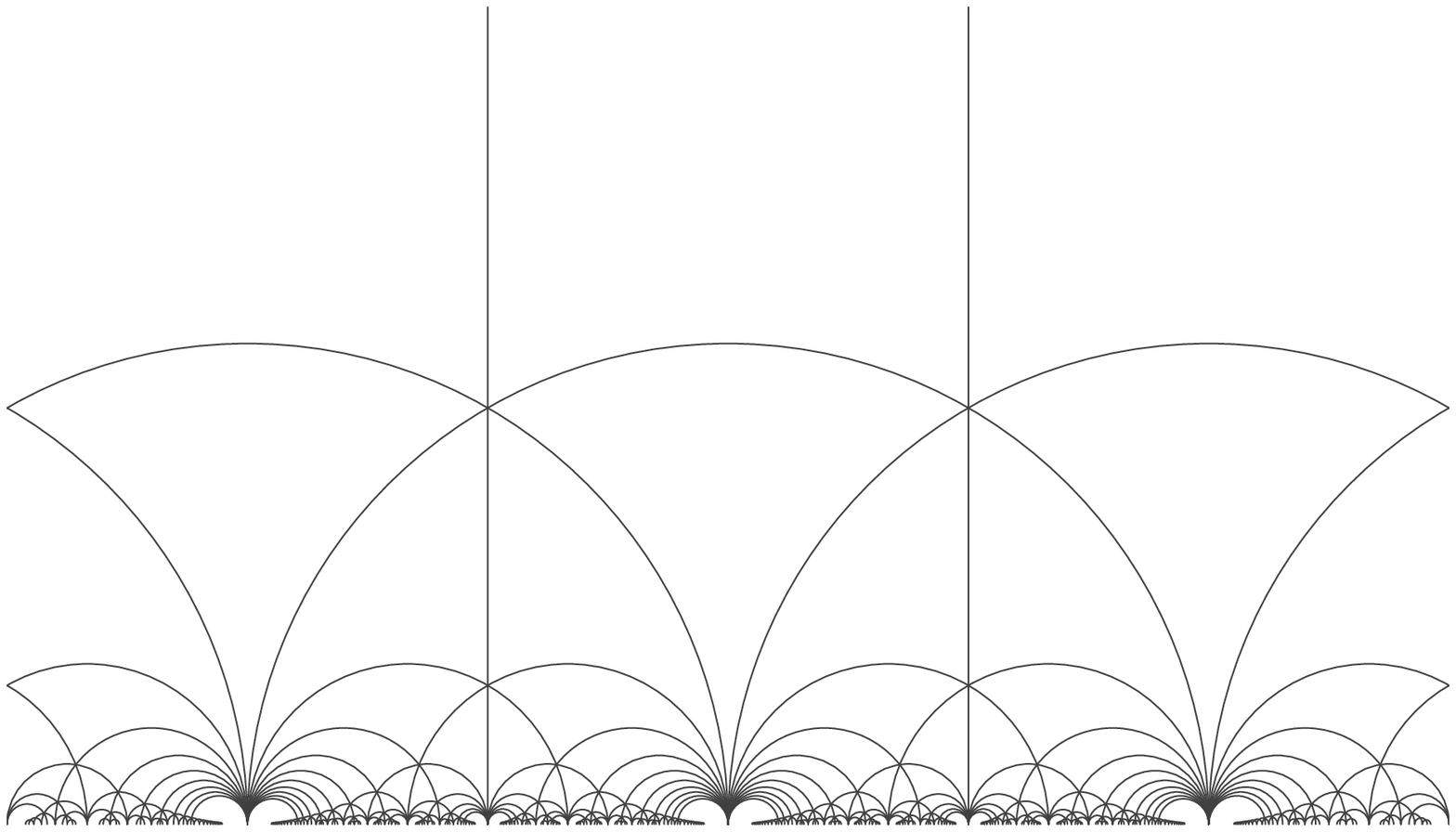}{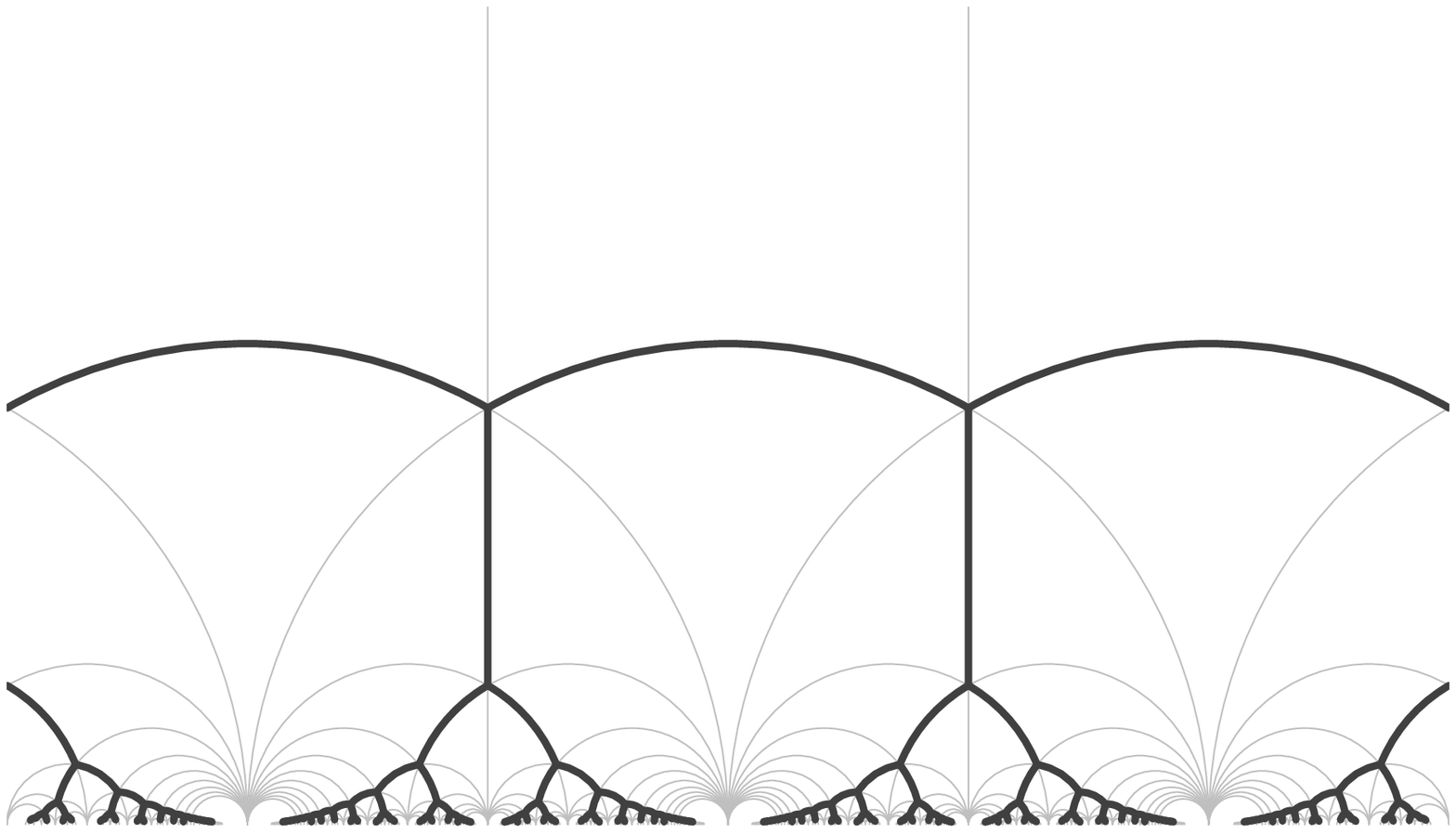}{5cm}{3.5cm}

It is believed that the phase structure as a function of $\tau$ of
a wide range of three-dimensional gravity theories (with different
sets of fields)
 is as we have just described.  For any
finite value of the dimensionless ratio $k=\ell/16 G$, the partition
function is smooth as a function of $\tau$, but for $k\to\infty$ or
$G\to  0$, it becomes non-smooth along the curves drawn in Fig. 3b.

We will address here  a puzzle that arises if one assumes
holomorphic factorization.  In this case, we take $k$ to be a
positive integer.  The partition function factorizes as $\hat
Z=Z_k(q)Z_k(\bar q)$, where the function $Z_k$ is a holomorphic and
modular-invariant function. $Z_k$ has an expansion near $q=0$
involving the negative energy Brown-Henneaux states,
\eqn\iko{Z_k=q^{-k}+\dots,} and has no other singularities.  In
order for $\hat Z$ to exhibit the phase structure sketched in Fig.
3b in the semiclassical limit, $Z_k$ must do the same.

The poses a puzzle, since $Z_k$ is holomorphic.  There is no
problem for a sequence of smooth but not holomorphic functions
depending on a parameter to become non-smooth along a phase
boundary in some limit. But how can this happen to a sequence of
holomorphic functions?

This question was actually answered by Lee and Yang
\refs{\leeyang,\fisher}. The original idea of Lee and Yang is that
although a system in finite volume can have no phase transition,
its partition function, in its dependence on the complexified
thermodynamic variables, can have zeroes.  Then, in the infinite
volume limit, the zeroes become more numerous and may become dense
or ``condense'' along a certain arc, and a true phase transition
can emerge.

In our problem, the limit $k\to \infty$ is analogous to a
thermodynamic limit, since the effective size of AdS space
(accessible at a given temperature) grows with $k$. In a
fundamental domain compactified by adding its ``cusp,'' the
function $Z_k$ has precisely $k$ poles, exhibited in \iko.  Hence
the number of zeroes in a fundamental domain is also equal to $k$.
Here in the case of zeroes that are on the boundary of the
fundamental domain, one must count only half the zeroes (or one
half of the boundary) to avoid double-counting.

Before going into details, we will give an informal explanation in
which we ignore the one-loop correction.  We consider the phase
boundary along the arc $|\tau|=1$, $|{\rm Re}\,\tau|\leq 1/2$
separating the thermal AdS phase from the Euclidean black hole. We will denote the portion of this arc with ${\rm Re}~\tau\ge0$ by $C$:
\eqn\asd{C = \{ \tau = x+ iy \big| |\tau|=1, 0 \le x \le 1/2
\}.} The other portion of the arc, with ${\rm Re}~\tau\le0$, is the image of $C$ under the modular transformation $\tau\to-1/\tau$. Now, the
contributions of $M_{0,1}$ and $M_{1,0}$ to the holomorphic
Poincar\'e series (ignoring the one-loop correction) are $\exp(-2\pi
i k \tau)$ and $\exp(-2\pi ik(-1/\tau))$.  On the arc $C$, we
have $1/\tau=\bar\tau$ and the sum of these two contributions is
\eqn\yto{\exp(-2\pi ik \tau)+\exp(2\pi i k\bar\tau).} The sum
vanishes if $\exp(2\pi i k (\tau+\bar\tau))=-1$.  Setting
$\tau=\exp(i\theta)$, $\pi/3\leq \theta\leq 2\pi/3$, we need
$\exp(4\pi i k \cos\theta)=-1$ or $4\pi k\cos\theta= (2n+1)\pi$,
$n\in\Z$.
On the arc $C$, we have $0\leq \cos\theta \leq 1/2$.  So this
equation has $k$ roots. In the limit $k\to\infty$, these roots
become dense everywhere along the arc.  This is the condensation of
Lee-Yang zeroes of the partition function.

Now we give a more precise account including the one-loop
correction.  However, the result does not depend on all the details
of the function $Z_k$. We will show, following \refs{\Rankin,\AKN},
that for a certain class of holomorphic functions, all zeroes lie on
the arc $C$ or one of its images under $SL(2,{\Z})$, as shown in Fig. 3.
In the large $k$ limit, the zeroes  become dense on $C$.

We consider a modular-invariant function $Z_{k}$ that is regular except
for a pole of order $k$ at $q=0$.  Such a function has  a Laurent
expansion \eqn\zform{\eqalign{Z_{k} (\tau) &= \sum_{\Delta=-k}^\infty
F_{\Delta} q^\Delta}.} We require that $F_{-k}=1$ and that the
coefficients $F_{\Delta}$, $\Delta\leq 0$, do not increase too
quickly. More precisely, we will show  that if \eqn\fbd{F_{\Delta} <
e^{2\pi (0.61)(\Delta + k)},~~~~~{\rm for}~\Delta \le0 } then the
zeroes are all as just described. The origin of the numerical
coefficient will be described below. This bound has a simple
physical explanation: when there are too many primary fields of low
dimension, i.e. too much light matter in the theory, the formation
of stable black holes is not possible. Problems with black hole
physics when there are many light species of matter have been
considered in, for example, \BekensteinJP.

We should emphasize that the extremal partition functions proposed in \WittenKT\ obey the bound \fbd.  So in the large $k$ limit the extremal partition functions exhibit the condensation of zeroes described above.  In fact,
for $\Delta<0$ and large $k$ the coefficients of the extremal partition function
grow much more slowly than \zform.   For large $\Delta+k$, they grow like
\eqn\asd{ F_{\Delta} \sim e^{{\pi\over 6}\sqrt{\Delta+k}} <<
e^{2\pi(0.61)(\Delta+k)}.}So one could add many more additional primaries with dimension $\Delta<0$ without spoiling the phase transition described above.

Our proof of the foregoing assertions  relies on a few specific
facts about modular functions, which we now review.

\subsec{Review: Properties of $J$}

First, we recall a few properties of the modular invariant
$J(\tau)$. Proofs of many of our assertions in this and the next
section can be found in \Apostol. The $J$ function has a
$q-$expansion \eqn\jdef{J(\tau) = \sum_{m\ge-1} c(m) q^m = {1\over
q} + 196884 q + \dots} where the coefficients $c(m)$ are positive
integers. Our choices of overall normalization and constant term
$c(0)=0$ agree with \flm\ but differ from the most classical
convention. As the $c(m)$ are positive, it follows that $J$ obeys
 \eqn\jsymmetry{J(\tau) = {\bar J} (-{\bar \tau})} and so
is real along the imaginary $\tau$ axis. In particular, it can be
shown that $J$ decreases monotonically along the imaginary axis
from $J(i \infty) = \infty$ to a minimum at  $\tau=i$ with $J(i) =
984$. (Continuing past $i$ on the imaginary axis, $J$ grows again
in view of its invariance under $\tau\to-1/\tau$.)

The modular domain \eqn\moddom{D=\{ \tau=x+iy \big| |\tau|\ge 1, -1/2
\le x \le 1/2 \} }is a fundamental region for $SL(2,{\Z})$.  On the
interior of $D$, $J$ takes every complex value exactly once. (This
follows from the fact that in the compactified fundamental domain,
$J$ has precisely one pole, which is at $\tau=i\infty$.) Using the
invariance of $J$ under $SL(2,{\Z})$ and \jsymmetry, one can show
that $J(\tau)$ takes real values on the boundary $\partial D$. In
particular, it turns out that on the arc $\tau = e^{i \theta}$,
$J(\tau)$ decreases monotonically from $984$ to $-744$ as $\theta$
runs from $\pi/2$ to $\pi/3$.  This is the arc we called $C$ above.  On the arcs $x=\pm 1/2$, $J$
decreases monotonically from $-744$ to $-\infty$ as $y$ runs from
$\sqrt{3}/2$ to $\infty$.

Although the exact form of $J(q)$ is quite complicated, in many
cases the ``tree-level approximation'' \eqn\jtree{J \sim q^{-1} +
\dots} is very useful. To this end, we note that for any given value
of $\tau=x+iy$, we have the bound \eqn\japprox{|J(\tau) - q^{-1} | =
|\sum_{m\ge 1} c(m) q^m| \le \sum_{m\ge1}c(m)|q|^m = J(iy) - e^{2\pi
y} .}  The function $ \sum_{m\ge1}c(m)|q|^m $ depends only on $y$
and is a monotonically decreasing function of $y$, so in the
fundamental domain $D$, it is bounded above by its value at
$y=\sqrt{3}/2$, which is the minimum of $y $ in $D$.  This value is
\eqn\asd{M = J(i\sqrt{3}/2) - e^{\pi \sqrt{3}}\approx 1335 .} So in
$D$, we have  \eqn\japproxx{ |J(\tau) - q^{-1}| \le M.}  Applying
the triangle inequality, it follows that, throughout $D$, we can
bound the value of $J$ by \eqn\jbound{ |J(\tau)| \le e^{2\pi y}+M .
}

The approximation \jtree\ is a good one when the imaginary part
of $\tau$ is large, i.e. when $\tau$ is close to the cusp at
$\tau\to\infty$.  By considering modular transformations of the
equation \japproxx\ we arrive at other approximations which are good
in other regions of the upper half plane. In particular, for any
element $\gamma\in SL(2,\Z)$, if $\gamma\tau$ is in the fundamental
domain $D$ we have \eqn\japproxxx{ |J(\tau) - e^{-2\pi i \gamma\tau}|
\le M,} and therefore \eqn\jboundd{ |J(\tau)| \le e^{2\pi {\rm
Im}~\gamma\tau}+M .} This final bound is particularly useful, for
the following reason. For a given value of $\tau$, consider the set
of possible values of ${\rm Im}~\gamma\tau$, for all $\gamma\in
SL(2,\Z)$.  This set has  a finite maximum value which occurs when
$\gamma\tau$ is in the fundamental domain or one of its images under
the map $\tau\to\tau+n$. So, using the formula \basd\ for ${\rm
Im}~\gamma\tau$, we conclude that for any value of $\tau=x+iy$ in
the upper half plane we have the bound \eqn\jbounddd{ |J(\tau)|\le
\exp\left\{{\rm max}_{c,d} {2\pi\,{\rm Im}~\tau\over
|c\tau+d|^{2}}\right\} +M .} Here the maximum is taken over all
relatively prime $c$ and $d$.

\subsec{Properties of $T_n J$}

We will now consider the action of the Hecke operator $T_n$ on $J$, defined by
\eqn\Tdef{T_n J = \sum_{\delta|n}\sum_{\beta=0}^{\delta-1} J({(n \tau + \beta \delta) /
\delta^2}).} This is a new modular invariant function with an $n$-th
order pole at $q=\infty$. Our normalization (which differs by a
factor of $n$ from that in the literature) has been chosen so that
the $q-$expansion of $T_n J$ starts with $q^{-n}$: \eqn\asd{T_n J =
\sum_{m=-n}^\infty c_n(m) q^m = q^{-n} + {\cal O}(q).} The Fourier
coefficients $c_n(m)$ of $T_n J$ are related to the coefficients
$c(m)$ of $J$ by \eqn\ad{c_n(m) = \sum_{\delta|(m,n)} {n\over \delta}
c\left({m n /\delta^2}\right)q^m .} From this it follows that the Fourier
coefficients of $T_n J$ are positive integers, just like those of
$J$.

From these facts, it is straightforward to see that $T_n J$
satisfies many of the same properties as $J$.  In particular, $T_n
J$ is real along the imaginary axis, as well as on the boundary of
the fundamental domain $\partial D$.  Moreover, just as with $J$,
the ``tree-level approximation'' \eqn\asd{T_n J\sim q^{-n} + \dots}
is good enough for many purposes.  This approximation is very good when $\tau\to i\infty$, but actually it can be extended to
understand the zeroes of $T_n J$, as was shown in \AKN. We will
summarize their argument here. In section 6.3, we apply the results
derived below to more general modular-invariant partition functions.

The basic idea is to apply either \japproxxx\ or \jbounddd\ to each
term in the sum \Tdef. This allows us to bound the value of
$T_{n}J(\sigma)$ for a point $\sigma=x+iy$ on the arc $C$ defined by
$|\sigma|=1$ and $0\leq x\leq 1/2$. Let us first consider the
$\delta=1$ term in \Tdef, which is $J(n\sigma)$. For any $\sigma$ in
$C$, we can find an integer $m$ such that $n\sigma+m$ lies in the
fundamental domain. This allows us to apply \japproxxx\ with
$\tau=n\sigma$ and $\gamma\tau=n\sigma+m$ to get \eqn\asd{
|J(n\sigma) -e^{-2\pi i n \sigma}| \le M .} Now, consider the
$\delta=n$, $\beta=0$ term in \Tdef, which is $J(\sigma/n)$. Since
$\sigma$ lies in $C$, we can find an integer $m$ such that
$-n/\sigma +m$ lies in the fundamental domain.  So we may apply
\japproxxx, with $\tau=\sigma/n$ and $\gamma=-n/\sigma +m$ to get
\eqn\asd{|J(\sigma/n) -e^{-2\pi i n/\sigma}| \le M.} Using these two
equations, we may apply the triangle inequality to \Tdef\ to get
\eqn\Tbd{ |T_{n}J(\sigma) - e^{-2\pi i n \sigma}- e^{-2\pi i
n/\sigma}| \le 2M + \sum{}'\left|J\left(n\sigma+\beta\delta\over
\delta^{2}\right)\right| .} Here the sum $\sum{}'$ is over the same
set of $\beta$ and $\delta$ as in \Tdef, except we have dropped the
two terms with $\delta=1$ and $\delta=n, \beta=0$.  We will now
apply \jbounddd\ to each term on the right hand side of \Tbd.
Consider first the term with $\delta=n$, $\beta=n-1$, which is
$J\left(\sigma-1\over n\right)$. We would like to apply \jbounddd\
with $\tau= {\sigma -1\over n}$, so we must ask what possible values
$|c\tau+d|$ can take for this value of $\tau$.  Now, since $\sigma$
lies on $C$, it follows that $|\sigma-1|>1$ and hence $|\tau| =
|{\sigma-1\over n}|>1/n $.  This implies that $|c\tau+d|\ge 1/n$ for
all possible choices of $c$ and $d$.  Since ${\rm Im}~\tau=y/n$,
equation \jbounddd\ gives \eqn\asd{ \left|J\left({\sigma + n-1\over
n}\right) \right| \le e^{2\pi n y} +M .}Let us consider the case
where $\delta=n$ and $0<\beta<n-1$, where we may apply \jbounddd\
with $\tau = {\sigma+\beta\over n}$.  For this range of $\beta$,
$|\sigma+\beta|>\sqrt{2}$, so $|\tau| > \sqrt{2}/n$.  Hence
$|c\tau+d|>\sqrt{2}/n$ and \japproxxx\ gives \eqn\asd{
\left|J\left({\sigma+\beta \over n}\right)\right| \le e^{\pi n y}+M
.}Finally, we consider the cases where $1<\delta<n$.  In this case
we apply \jbounddd\ with $\tau=(n\sigma + \beta\delta)/\delta^{2}$.
The fact that  $\sigma$ lies on $C$ implies that
$|c\tau+d|>\sqrt{3}n/\delta^{2}$.  So we end up with the bound
\eqn\asd{ \left|J\left({\sigma+\beta \over n}\right)\right| \le
e^{{2\pi n\over 3} y}+M .}

Putting this all together, equation \Tbd\ becomes \eqn\tineq{
|T_{n}J(\sigma) - e^{-2\pi i n \sigma}- e^{-2\pi i n/\sigma}|
\le e^{2\pi n y} + n e^{\pi n y}+n^{2}e^{{2\pi n\over 3}y} + n^{2} M
.} The factors of $n$ and $n^{2}$ in \tineq\ come from the simple fact that
there are less than $n$ terms with $\delta=n$, and less than $n^{2}$ terms with $1<\delta<n$.
 Multiplying both sides of \tineq\ by $e^{-2\pi n y}$, and using the fact that
 $e^{-2\pi in/\sigma} = e^{2\pi i n\bar \sigma}$ for points on the curve $C$, this becomes
\eqn\tineqq{|T_n J e^{-2\pi n y} - 2\cos(2\pi n x)| \le 1 + n
e^{-\pi n y} + n^{2}e^{-{4\pi n\over 3}y}+ M n^{2}e^{-2\pi n y}.}
For the moment, we concentrate on the case $n\geq 2$. Since $y\ge
\sqrt{3}/2$ for any point on $C$, we may evaluate this right hand
side of \tineqq\ to get \eqn\Tbdd{|T_n J e^{-2\pi n y} - 2\cos(2\pi
n x)| \le 1.12 .} This formula is valid for any point
 on the arc $C$ and $n\ge2$.

Equation \Tbdd\ is an important result.  It places strong
constraints on the location of the zeros of $T_{n}J$. To see this,
note that on each of the $n$ intervals $I_m$ defined by ${m\over n}
< x < {m+1\over n}$, $m=0,\dots,n-1$, the function $2 \cos(2\pi n
x)$ varies monotonically from $-2$ to $2$. The right hand side of
\Tbdd, however, is less than $2$. This implies that $T_n J$ must
have $n$ distinct zeroes on $C$, with one in each interval $I_m$. As
$n\to\infty$, the lengths of these intervals vanish and the zeroes
of $T_n J$ become dense on $C$. Finally, note that, as $T_n J$ is an
$n^{th}$ order polynomial in $J$, it has only $n$ zeroes on $D$. In
particular, it has no additional zeroes beyond those described
above.

We have only proven the bound \Tbdd\ for $n\ge2$. For our
application, we also need the $n=0$ and $n=1$ cases, which we will
consider separately. For $n=0$, we define $T_{0}J=1$. So the bound
\Tbdd\ is trivial. For $n=1$, we have $T_{1}J = J$. In this case we
have the slightly weaker bound\eqn\jbd{ |J(\tau) e^{-2\pi y} -2
\cos(2\pi x)| \le 1.22 } for points on $C$.  This is straightforward
to verify numerically, although it may also be understood
analytically.  The point is that the function $J(\tau) e^{-2\pi y} -
2 \cos(2\pi x)$ is monotonic along the arc $C$.  So the value of
this function on $C$ is bounded by its values at the endpoints
$\tau=i$ and $\tau=1/2 + \sqrt{3}/2 i$.  Using $J(i)= 984$ and
$J(1/2+\sqrt{3}/2 i)=-744$ we may compute the values of the function
at these endpoints, giving the bound \jbd.

\subsec{Zeroes of Modular Invariant Partition Functions}

We are now ready for the general case of a modular invariant
partition function at level $k$: \eqn\asd{Z_k (\tau) =
\sum_{\Delta=-k}^\infty F_{\Delta} q^{\Delta} = \sum_{\Delta=-k}^0
F_{\Delta} T_{|\Delta|} J(\tau)} where the $F_{\Delta}$ are
non-negative integers and $F_{-k}=1$.  As
with $J(\tau)$, the fact that $Z_k(\tau)$ has real coefficients and
is modular-invariant means that it is real on the imaginary $\tau$
axis as well as on the boundary of the fundamental region $\partial
D$.

In the previous section, we proved the estimate \Tbdd\ for
$T_{n}J$ on $C$ with $n\ge2$.  Along with the estimate \jbd\ on $C$, this implies that for all $n\ge0$ \eqn\pasd{T_n J = e^{2\pi n y} ( 2 \cos(2\pi n x) +
E_n)} where $E_{n}$ is an error term obeying\eqn\asd{|E_n| < 1.22 .}
By simply adding up the inequalities \pasd\
for $n=0,\dots,k$, with coefficients $F_{-n}$, and using the fact
that $F_{-k}=1$, we get a  a similar estimate for $Z_k$:
\eqn\qasd{
 Z_k e^{-2 \pi k y}-2 \cos(2 k \pi x)
= E_k+\sum_{\Delta=-k+1}^{0} F_{\Delta} e^{-2\pi (k+\Delta)y}\left(2
\cos(2 \pi \Delta x) +E_{|\Delta|}\right).}

To bound the location of the zeroes of $Z_{k}$, we must show that
the right hand side is less than $2$. Then, as in the previous
section, the zeroes of $Z_{k}$ will lie on $C$ and become dense in
the large $k$ limit.

For this to be the case, $F_{\Delta}$ must not grow too quickly with
$\Delta$.  For example, assume that \eqn\wasd{ F_{\Delta} < A e^{2
\pi \alpha(\Delta + k)}~~~~~{\rm for}~-k\le\Delta\le0} where
$\alpha$ and $A$ are positive constants.  In this case
\eqn\casd{\eqalign{ \sum_{\Delta=-k+1}^{0} F_{\Delta} e^{-2\pi
(k+\Delta)y} &< A \sum_{\Delta=-k+1}^{0}
e^{2\pi(k+\Delta)(\alpha-y)}
\cr
&< A\sum_{n=1}^{k} e^{-2\pi n(\alpha-\sqrt{3}/2)}
<
{A\over e^{2\pi (\alpha-\sqrt{3}/2)}-1} .}} In the second line, we
have used the fact that $y>\sqrt{3}/2$ on $C$. Since
\eqn\basd{|2\cos (2\pi \Delta x) + E_{|\Delta|}| \le 2 + |E_{|\Delta|}|
<3.22,} it follows from \asd\ that \eqn\asd{| Z_k e^{-2 \pi k y}-2
\cos(2 k \pi x)| \le |E_{k}| + 3.22{ A\over
e^{2\pi(\sqrt{3}/2-\alpha)}-1},}which is less than $2$ for certain
values of $A$ and $\alpha$.  In particular, using $|E_{k}|<1.22$ and
setting $A=1$, we find that the right hand side is less than $2$
provided that \eqn\asd{ \alpha \le 0.61 } We have used several
rather conservative estimates, so it is quite possible that the
zeroes of $Z_{k}$ condense on $C$ even when the coefficients
$F_{\Delta}$ do not satisfy the precise bound given above.

It is easy to check that this bound is indeed satisfied for the
extremal partition functions of \WittenKT.  In fact, the
coefficients $Z_k(\Delta)$ grow much more slowly than the required
behavior, indicating that many more primary fields could be added
without spoiling the existence and nature of the phase transition.

\newsec{Supergravity Partition Functions}

Our goal here will be to repeat the analysis of Secs. 2 and 3 for
supergravity.  We will consider only the basic case of ${\cal N}=1$
supergravity, in which the symmetry group $SL(2,\R)\times SL(2,\R)$
of $\AdS_3$ is replaced by $OSp(1|2)\times OSp(1|2)$, where
$OSp(1|2)$ is a supergroup whose bosonic part is
$Sp(2,\R)=SL(2,\R)$. The boundary CFT has $(1,1)$ supersymmetry,
that is, ${\cal N}=1$ supersymmetry for both left- and right-movers.

\subsec{Formalism}

In supergravity, there are a few closely related choices of possible
partition function.  We can compute either $\Tr\,\exp(-\beta H-i\theta J)$ or
$\Tr\,(-1)^F\exp(-\beta H-i\theta J)$.  And we can compute this trace in
either the Neveu-Schwarz (NS) or the Ramond (R) sector.  As usual,
we attempt to compute these partition functions  by summing over
three-manifolds $M$ that are locally $\AdS_3$ and whose conformal
boundary $\Sigma$ is a Riemann surface of genus 1. The four possible
partition functions (NS or R, with or without an insertion of
$(-1)^F$) correspond to the four spin structures\foot{In fact, using
a discrete $R$-symmetry that appears to be present in the boundary
CFT, we could choose different spin structures for left- and
right-movers on $\Sigma$, but we will not consider this
generalization.} on $\Sigma$.  The three that correspond to odd spin
structures are permuted by the action of $SL(2,\Z)$, as we further
discuss below.

Once we pick the spin structure on $\Sigma$, we then sum over
choices of $M$ such that the given spin structure on $\Sigma$ does
extend over $M$.  For example, what spin structure on $\Sigma$ is
compatible with taking $M=M_{0,1}$ to be the three-manifold
related to perturbative excitations of $\AdS_3$?

In $M_{0,1}$, the ``spatial'' circle on $\Sigma$ is contractible.
This means that the NS spin structure on the spatial circle
extends over $M_{0,1}$ and the R spin structure does not.  Hence
$M_{0,1}$ contributes to traces in the NS sector, not the R
sector.

What is the spectrum of thermal excitations in the NS sector? In
ordinary gravity, the thermal excitations of left-movers are
obtained by acting on the ground state $|\Omega\rangle$ with
Virasoro generators $L_{-n}$, $n\geq 2$.  When the boundary theory
has ${\cal N}=1$ supersymmetry, we can also act on
$|\Omega\rangle$ with superconformal generators $G_{-n+1/2}$,
$n\geq 2$.  (We recall that $G_{-1/2}|\Omega\rangle=0$.) Writing
$-k^*/2$ for the ground state energy, the partition function  of
left-moving excitations is therefore
\eqn\tomixo{q^{-k^*/2}\prod_{n=2}^\infty{1+q^{n-1/2}\over 1-q^n}.}
Including both left- and right-moving excitations, the
contribution of $M_{0,1}$ to $F(q,\bar q)=\Tr_{\rm
NS}\,\exp(-\beta H-i\theta J)$ is
\eqn\omixo{F_{0,1}=\left|q^{-k^*/2}\prod_{n=2}^\infty{1+q^{n-1/2}\over
1-q^n}\right|^2.}  This formula can be justified exactly as
in Sec. 2.2 for the bosonic case.

The complete function $F$, or more precisely the sum of known
contributions to it, is evaluated by summing over all those modular
images of $M_{0,1}$ over which the relevant spin structure extends.
We can represent the four spin structures on the two-torus $\Sigma$
by a column vector \eqn\yolpo{\left(\matrix{\mu\cr\nu\cr}\right),}
where $\mu$ and $\nu$ represent respectively the fermion boundary
conditions in the ``time'' and ``space'' directions on the two-torus
$\Sigma$. We consider $\mu$ and $\nu$ to be valued in $\half\Z/\Z$,
taking the values  1/2 for antiperiodic (NS) boundary conditions and
0 for periodic (R) ones. An element of $SL(2,\Z)$ acts on the spin
structures by
\eqn\zopox{\left(\matrix{\mu\cr\nu\cr}\right)\to\left(\matrix{a&b\cr
c&d\cr}\right)\left(\matrix{\mu\cr\nu\cr}\right).}

\def\odd{{\rm odd}}
The NS partition function $F(\tau)=\Tr_{\rm NS}\,\exp(-\beta
H-i\theta J)$ corresponds to $\mu=\nu=1/2$.  This condition is
invariant under the subgroup of $SL(2,\Z)$ characterized by saying
that $c+d$ and $a+b$ are both odd. If $c+d$ is odd, we can make
$a+b$ odd by adding to $(a,b)$ a multiple of $(c,d)$.  $F$, or at
least the sum of known contributions to it, can therefore be
computed by summing $F_{0,1}$ over modular images with $c+d$ odd:
\eqn\telox{F(\tau)=\sum_{c,d|c+d\,\odd}F_{c,d}(\tau)} or
equivalently
\eqn\zelox{F(\tau)=\sum_{c,d|c+d\,\odd}F_{0,1}((a\tau+b)/(c\tau+d)).}
For any given $c,d$ such that $c+d$ is odd, we pick $a$ and $b$
such that $ad-bc=1$ and $a+b$ is odd.  Because $F_{0,1}(\tau)$ is
invariant under $\tau\to\tau+2$, the summand in \zelox\ does not
depend on this choice.

It is also of interest to compute partition functions with other
spin structures.  However, the three even spin structures on
$\Sigma$ (the ones for which $\mu$ and $\nu$ are not both zero)
are permuted by $SL(2,\Z)$ and so the associated partition
functions are not really independent functions. If we set $\mu=0$,
$\nu=1/2$, we get $ G(\tau)=\Tr_{\rm NS}\,(-1)^F\exp(-\beta
H-i\theta J)$. The contribution of $M_{0,1}$ to this partition
function is obtained by reversing the sign of all fermionic
contributions in \omixo: \eqn\zomixo{G_{0,1}(\tau)=
\left|q^{-k^*/2}\prod_{n=2}^\infty{1-q^{n-1/2}\over
1-q^n}\right|^2.} The subgroup of $SL(2,\Z)$ that preserves the
conditions $\mu=0$, $\nu=1/2$ is characterized by requiring that
$b$ should be even, which implies that $a$ and $d$ are odd.  Hence
\eqn\ombixo{G(\tau)=\sum_{c,d|d\,\odd} G_{c,d}=\sum_{c,d|d\,\odd}
G_{0,1}((a\tau+b)/(c\tau+d)),} where for given $c,d$, we pick
$a,b$ so that $ad-bc=1$ and $b$ is even.  A modular transformation
$T:\tau\to\tau+1$ exchanges $(\mu,\nu)=(0,1/2)$ with
$(\mu,\nu)=(1/2,1/2)$, so in particular
\eqn\bilxo{F(\tau)=G(\tau+1)=F(\tau+2),} and  the summand in
\ombixo\ does not depend on the choice of $a,b$.

The Ramond partition function $K=\Tr_{\rm R}\exp(-\beta H-i\theta
J)$ is computed from $\mu=1/2$, $\nu=0$, so
\eqn\lugl{K(\tau)=G(-1/\tau).} This completes our characterization
of three of the four partition functions. Finally, in any
supersymmetric theory with discrete spectrum, the fourth partition
function $I=\Tr_{\rm R}(-1)^F\exp(-\beta H-i\theta J)$ is an
integer, independent of $\beta$ and $\theta$, since it can be
interpreted as the index of a supersymmetry generator. It must be
computed using the odd spin structure, the one with $\mu=\nu=0$. In
three-dimensional gravity, assuming that the partition function can
be computed by summing over smooth three-geometries, $I$ vanishes,
since the odd spin structure does not extend over any three-manifold
with boundary $\Sigma$.

\bigskip\noindent{\it General Structure}

Some simple but important facts follow from \bilxo.  Since
$F(\tau)=\Tr_{\rm NS}\,\exp(-\beta H-i\theta J)$ is invariant under
$\tau\to \tau+2$, it follows that all eigenvalues of $J=L_0-\bar
L_0$ take values in $\Z/2$.  The transformation $\tau\to\tau+1$
acts as 1 or $-1$ on states with integer or half-integer $J$; thus
it acts as $(-1)^{2J}$.  On the other hand, if we insert a factor
of $(-1)^F$ in the trace, we get $G(\tau)=\Tr_{\rm
NS}\,(-1)^F\exp(-\beta H-i\theta J)$.  Since $G(\tau)=F(\tau+1)$, the
operator $(-1)^F$ is equivalent to $(-1)^{2J}$.  In other words,
states of integer or half-integer $J$ are bosonic or fermionic,
respectively.  The exact theory (to the extent that it can be
reconstructed from the sum over smooth classical geometries)
inherits this property from the perturbative spectrum of
Brown-Henneaux excitations.

The general form of $G(\tau)$, assuming that it has a Hilbert
space interpretation, should therefore be \eqn\ofus{G(\tau)=
\sum_{j\in \Z}\sum_{n}a_{n,j}\exp(-\beta E_{n,j}+i\theta j)-\sum_{j\in
\Z+1/2}\sum_{n}a_{n,j}\exp(-\beta E_{n,j}+i\theta j).} Here $E_{n,j}$,
$n=1,2,3,\dots$, are the energy eigenvalues for states of angular
momentum $j$, and $a_{n,j}$ is the number of states of energy
$E_{n,j}$.  The $a_{n,j}$ are positive integers; the minus sign
preceding the second term in \ofus\ reflects the relation
$(-1)^{2J}=(-1)^F$.

\subsec{The Computation}

We we will now compute the partition functions of the previous
section by summing over smooth geometries.  We will consider the
partition function $G(\tau)=\Tr_{\rm NS}\,(-1)^F\exp(-\beta
H-i\theta J)$, as it is technically the simplest; the two other
non-zero partition functions are then given by \bilxo\ and \lugl.

From \zomixo\ and \ombixo\ we have \eqn\asd{
{G}(\tau)=\sum_{c,d|~d~\odd}
\left.\left(|q|^{-k^{*}}\prod_{n\ge2}\left|{1-q^{n-1/2}\over
1-q^{n}}\right|^{2}\right)\right|_{\gamma} .} To understand the modular
transformation properties of this sum, it is useful to rewrite the
infinite product in terms of Dedekind eta functions. Using the
identities \eqn\asd{ \prod_{n=2}^{\infty}(1-q^{n})^{-1} =
{q^{1/24}(1-q)\over \eta(\tau)}} and \eqn\asd{
\prod_{n=2}^{\infty}(1-q^{n-1/2}) = {q^{1/48} \over
(1-q^{1/2})}{\eta(\tau/2)\over \eta(\tau)} } this may be written as
\eqn\asd{ G(\tau) = \sum\left.\left(|q|^{-k^{*} +3/24} |1+q^{1/2}|^{2}
{|\eta(\tau/2)|^{2}\over |\eta(\tau)|^{4}}\right)\right|_{\gamma}. }

We may now extract these Dedekind eta functions from the sum, using
the fact that $\sqrt{{\rm Im} ~\tau}~|\eta(\tau)|^{2}$ is modular
invariant: \eqn\asd{\eqalign{ {G}(\tau) &= {|\eta(\tau/2)|^{2}\over
y^{1/2}|\eta(\tau)|^{4}} \sum  \left(y^{1/2} |q|^{-k^{*}+3/24}
|1-q^{1/2}|^{2}\right)_{\gamma}\cr &= {|\eta(\tau/2)|^{2}\over
y^{1/2}|\eta(\tau)|^{4}} \big({\hat E}( k^{*}-3/24,0)+{\hat
E}(k^{*}+1-3/24,0)+\cr&~~~~~~~~~~~~~~~~{\hat E}(k^{*}+1/2-3/24,1/2)
+{\hat E}(k^{*}+1/2-3/24,-1/2)\big) }.} In the second line we have
defined \eqn\asd{ {\hat E}(\kappa,\mu) = \sum_{c,d|~d~\odd} {y^{1/2}
\over |c\tau + d|^{}}  \exp\left\{2\pi \kappa~ {\rm Im}~ \gamma \tau
+ 2\pi i\mu~{\rm Re}~ \gamma \tau\right\} .} This is the
supersymmetric version of the Poincar\'e series studied in Sec. 3.

This sum is divergent, for the same reasons described in Sec. 3.1.
In particular, at large $c$ and $d$ the exponential approaches one
and we are left with the linearly divergent sum $\sum_{c,d}
|c\tau+d|^{-1}$. The sum may be regularized by considering the more
general Poincar\'e series \eqn\edeff{ {\hat E}(s,\kappa,\mu) =
\sum_{c,d|~d~\odd} { y^{s} \over |c\tau + d|^{2s} } \exp\left\{2\pi
\kappa~ {\rm Im}~ \gamma \tau + 2\pi i\mu~{\rm Re}~ \gamma
\tau\right\}}as an analytic function of $s$ and taking $s\to1/2$.
This regularization scheme can be justified on physical grounds,
following the same line of argument presented in Sec. 3.1.

We will now proceed to
compute the Fourier coefficients
of the sum \edeff.  Our calculation
is quite similar to that done in Sec. 3.
The only differences are that in \edeff, $\kappa$ and $\mu$ are allowed to be
half-integer, and that we are summing over
$c$ and $d$ such that $(c,d)=1$ and $d$ is odd.
These two conditions on $c$ and $d$ can be combined in
to the single condition $(2c,d)=1$.

We start by letting $d=d'+2nc$, where $d'=d~{\rm mod}~2c$.  The sum
in \edeff\ can be written as a sum over $c,d'$, and $n$:
\eqn\bigsum{ E(s,\kappa,\mu) = y^{s}e^{2\pi (\kappa y+i \mu x)} +
\sum_{c>0}\,\sum_{d'\in\Z/2c\Z}\, \sum_{n\in{\Z}} f(c,d',n)}
where\eqn\fis{ f(c,d',n) ={y^s\over |c(\tau +2n) + d'|^{2s}}
\exp\left\{ {2\pi \kappa y \over | c(\tau+2n) +d'|^{2}} + 2\pi i
\mu~ \left({a\over c} - {cx+d\over c |c(\tau +2n) +d'|^{2}}\right)
\right\}. } We will now apply the Poisson summation formula to the
sum over $n$, as we did for the bosonic partition function in Sec.
3.2. First, we must compute the Fourier transform
\eqn\fint{\eqalign{ {\hat f} (c,d',{\hat n}) &=
\int_{-\infty}^{\infty} dn~ e^{2\pi i n {\hat n}} f(c,d',n)\cr
 &= {1\over 2}\exp\left({2\pi i {(2\mu) a-{\hat n}d'\over 2c}
- \pi i{\hat n } x}\right) \int_{-\infty}^{\infty} d{t}~ e^{\pi i
{\hat n} t} \left( {y\over c^{2}(t^{2}+y^{2})} \right)^{s}
\exp\left\{{ 2\pi (\kappa\,y -i\mu t)\over c^{2}(t^{2}+y^{2})}
\right\}.\cr}} We have written the integral in terms of a shifted
integration variable  $ t= 2n + x + {d'\over c}$.

Recall that, as described after equation \zomixo, we must chose our integers $a$ and $b$ such that $b$ is
even.  The determinant condition $ad-bc=1$ therefore implies that
$ad'=1~{\rm mod}~2c$.  Since $2\mu\in\Z$, this implies that the
exponential prefactor in \fint\ is a function only of $d'=d~{\rm
mod}~2c$.  We may therefore extract the $d'$ dependence of the sum
\edeff\ into the Kloosterman sum $S(-{\hat n},2\mu;2c)$, as defined
in \klooster.  The integral appearing in \fint\ is precisely that
defined in \idef. The Fourier cofficients of the Poincar\'e series
\edeff\ are therefore given by \eqn\asd{ {\hat E}(s,\kappa,\mu) =
y^{s}e^{2\pi (\kappa y+i \mu x)} + \sum_{\hat n} e^{\pi i {\hat n}x}
{\hat E}_{\hat n}(s,\kappa,\mu) } where \eqn\eis{ {\hat E}_{\hat
n}(s,\kappa,\mu) = {1\over 2} \sum_{m=0}^{\infty} I_{m,{\hat
n/2}}(s,\kappa,\mu)~ y^{1-m-s}~
\left(\sum_{c=1}^{\infty}c^{-2(m+s)}S(-{\hat n},2\mu;2c) \right) }
is defined in terms of the integrals \idef.

\def\even{{\rm even}}
We will now restrict our attention to the $\hat n=0$ case. First
consider $\mu=0$. In this case, the integrals were given in \izero.
To do the sum over $c$, note first that the Kloosterman sum
$S(0,0,2c)$ is equal to Euler's totient function $\phi(2c)$. The sum
over $c$ is \eqn\phievenn{\sum_{c>0} c^{-2(m+s)} S(0,0,2c) =
\sum_{c>0} c^{-2(m+s)} \phi(2c)= {2^{2(m+s)}\over2^{2(m+s)}-1}
{\zeta(2(m+s)-1)\over \zeta(2(m+s))} .} To prove this formula, we
first evaluate the sum $\sum_{c~{\rm odd}}c^{-\sigma}\phi(c)$, where
we have defined $\sigma=2(m+s)$.  Multiplying this sum by
\eqn\zetaodd{ \sum_{n~{\odd}}n^{-\sigma} =
(1-2^{-\sigma})\zeta(\sigma) } gives \eqn\asd{\eqalign{
(1-2^{-\sigma})\zeta(\sigma) \sum_{c~\odd}c^{-\sigma} \phi(c) &=
\sum_{c,n~\odd}(cn)^{-\sigma} \phi(c) \cr &= \sum_{m~\odd}
m^{-\sigma} \left(\sum_{c|m} \phi(c)\right) \cr &=\sum_{m~\odd}
m^{-\sigma+1} = (1-2^{1-\sigma})\zeta(\sigma-1) }.} In the second
line we defined the new variable $m=cn$ to be summed over, and in
the third line we have used the basic identity for the totient
function $ \sum_{c|m}\phi(c) = m$. We thus have \eqn\phiodd{
\sum_{c~\odd} c^{-\sigma}\phi(c) = {2^{\sigma}-2\over2^{\sigma}-1}
{\zeta(\sigma-1)\over \zeta(\sigma)} .}
 The sum over even $c$ may
be done by recalling the identity that was proved in  Sec. 3  by
similar methods: \eqn\asd{ \sum_{c>1}c^{-\sigma} \phi(\sigma) =
{\zeta(\sigma-1) \over \zeta(\sigma)} } to get \eqn\phieven{
\sum_{c~\even}c^{-\sigma}\phi(c) = {1\over2^{\sigma}-1}
{\zeta(\sigma-1)\over \zeta(\sigma)}.} But
\eqn\rex{\sum_{c~\even}c^{-\sigma}\phi(c)= 2^{-\sigma}
\sum_{c>0}c^{-\sigma}\phi(2c).} Combining the last two formulas,
we arrive at \phievenn.

As in the bosonic case, we must be careful when taking $s\to1/2$.
For $m=0$ (and $\hat n=0$), the sum \phievenn\ vanishes at $s=1/2$,
whereas the integral $I_{0,0}$ has a pole at $m=0$, $s=1/2$, as we
see in \izero. The product of the two factors is finite; in fact,
$\Gamma(s-1/2)/\zeta(2s)\to2$ at $s\to1/2$. The $m>0$ terms are
finite without any such subtleties. Evaluating the first three terms
in the expansion of \edeff\ gives \eqn\asd{ {\hat E}_0(1/2,\kappa,0)
= - y^{1/2} + \left({8 \pi^{3}\over 21
\zeta(3)}\kappa\right) y^{-1/2} + \left({64 \pi^{6}\over
4185\zeta(5)}\kappa^{2}\right)y^{-3/2} + {\cal O}(y^{-5/2}) .}

Let us now consider the $\mu=\pm1/2$ terms.  In this case the
integrals are only slightly different from those done in Sec. 3,
which were evaluated at $\mu=\pm1$.  For $m=0$, we find that
\eqn\ionezerosup{ I_{0,0}(s,\kappa,\pm1/2) = \sqrt{\pi}
{\Gamma(s-1/2)\over \Gamma(s)} .}This is the analog of equation
\ionezero\ used in the computation of the bosonic partition
function.  For $m>0$ the integrals are complicated hypergeometric
functions of the sort written down in \ione.  At $s=1/2$ these
hypergeometric functions simply to give \eqn\ionesup{
I_{m,0}(1/2,\kappa,\pm1/2) = {2^{1-m}\pi^{m+1/2}\over m
\Gamma(m+1/2) } T_{m}(2\kappa) .} This is the analog of equation
\ionesimp\ used in the bosonic case.

To do the sum over $c$, we use the fact that the Kloosterman sum
$S(0,\pm1,2c)$ is equal to the M\"{o}bius function $\mu(2c)$. The
sum over $c$ is given by \eqn\sssum{ \sum_{c>0}c^{-2(m+s)}
S(0,\pm1,2c) = \sum_{c}c^{-2(m+s)}\mu(2c) = -{2^{2(m+s)}\over
(2^{2(m+s)}-1)\zeta(2(m+s))}. } We can obtain this formula as
follows. We first compute the sum $\sum_{c~\odd}c^{-\sigma}\mu(c)$.
Multiplying this sum by \zetaodd\ we get \eqn\asd{\eqalign{
(1-2^{-\sigma})\zeta(\sigma) \sum_{c~\odd}c^{-\sigma} \mu(c) &=
\sum_{c,n~\odd}(cn)^{-\sigma} \mu(c) \cr &= \sum_{m~\odd}
m^{-\sigma} \left(\sum_{c|m} \mu(c)\right) \cr &=1 }.} In the second
line we have let $m=cn$ and in the third line we have used the fact
that for any $m$, $\sum_{c|m}\mu(c)=\delta_{m,1}$.
 The sum over even $c$ may
be done by recalling the similar identity from Sec. 3 \eqn\asd{
\sum_{c>1}c^{-\sigma} \mu(\sigma) = {1 \over \zeta(\sigma)} } to
get \eqn\mueven{ \sum_{c~\even}c^{-\sigma}\phi(c) =
-{1\over2^{\sigma}-1} {1\over \zeta(\sigma)},} which is equivalent
to \sssum.

At $s=1/2$, we again must be careful
to cancel the zero in \sssum\ at $m=0$ against
a pole in \ionezerosup, using the fact that $\Gamma(s-1/2)/\zeta(2s)\to2$ as $s\to1/2$.
Including the next two terms in the series, we find
\eqn\asd{
E_0(1/2,\kappa,\pm1/2) = -2 y^{1/2} -
\left({16 \pi^{}\over 7 \zeta(3)}\kappa\right) y^{-1/2}
 - \left({16 \pi^{2}\over 93\zeta(5)}
 (8\kappa^{2}-1)\right)y^{-3/2} + {\cal O}(y^{-5/2})
}

Putting this all together gives the following expansion for the
partition function \eqn\gis{\eqalign{ {G}(\tau) = &{G}_{0,1}(\tau)
+ {|\eta(\tau/2)|^{2}\over |\eta(\tau)|^{4}} \Big(-6
+{(6+16k^{*})(\pi^{3}-6\pi)\over 21 \zeta(3)}y^{-1}\cr & -
{4\pi^{2}(2880 k^{*}{}^{2}-16k^{*}(2\pi^{4}-135) +45-12\pi^{4}  )  \over 4185 \zeta(5)}y^{-2}
+{\cal O}(y^{-3})\Big) .}} As in the bosonic case, a Hilbert space
interpretation is precluded both by the fact that the coefficient
of the leading correction to $G_{0,1}(\tau)$ is negative and by
the fact that there are additional corrections involving powers of
$1/y$. A Hilbert space interpretation would require the leading
coefficient to be positive in view of \ofus; the leading
correction to the Brown-Henneaux spectrum arises for an integer
value of the angular momentum, namely $j=\hat n=0$.

We have so far considered just the $\hat n =0$ case.  The discussion
for the $\hat n\ne 0$ case will be qualitatively similar to the
bosonic case discussed in Sec. 3.4.  The integrals can be evaluated
explicitly -- they are the same ones we considered in Sec. 3.4 --
and contribute terms to the NS sector partition function which fall
off exponentially at large $y$.   As in the bosonic case, the sums
over $c$  cannot be expressed in terms of elementary number
theoretic functions but are related to an appropriately defined
Selberg zeta function. In particular, they should be related to the
Selberg zeta function associated to the congruence subgroup
described in Sec. 7.1, where $b$ is constrained to be even.  As in
Sec. 3.4, this should provide a finite regularization of the sum
because the associated Laplacian does not have a discrete
eigenfunction with eigenvalue $1/4$.

We conclude this section by emphasizing a few possible
interpretations of the result \gis\ for the supergravity partition
function $G(\tau)$. As in the bosonic case, this partition function
does not have the structure of a Hilbert space trace ${\rm Tr}_{\rm
NS}~(-1)^{F}e^{-\beta H - i\theta J}$. One possible interpretation
of this result is that three-dimensional pure supergravity does not
exist as a quantum theory, and that additional degrees of freedom
must be included in order to render the theory sensible. A second
possible implication is that the pure supergravity theory exists,
but contains long strings, similar to those described in Sec. 4.1.
A third possibility is that additional complex saddle points must be
included in the sum over geometries, as described for the bosonic
theory in Sec. 4.2.  This might lead to a sum over two copies of the
modular group, resulting in a holomorphic partition function which
might coincide with the extremal supergravity partition functions
described in  \WittenKT.

\subsec{Phase Transitions}

In this section we will comment briefly on the supergravity
generalization of the results of Sec. 6. In particular, we ask
whether, for holomorphically factorized supergravity partition
functions, Hawking-Page phase transitions occur via a condensation
of zeroes along curves in the complex temperature plane. Numerical
evidence described to us by M. Kaneko suggests that this may be the
case.

Let us first ask what the phase diagram of three-dimensional
supergravity should be.  We will consider the $NS$ partition
function $F(\tau) = {\rm Tr}_{\rm NS}~e^{-\beta H - i\theta J}$.  As
described in Sec. 7.1, $F(\tau)$ is invariant under the subgroup of
$SL(2,\Z)$ defined by the condition that $a+b$ and $c+d$ are both
odd; this group is sometimes called $\Gamma_{\theta}$.  The
tessellation of the upper half plane ${\eusm H}$ by fundamental
domains of $\Gamma_{\theta}$ is shown in Fig. 4a).

\doublefig{ a) The tessellation of the upper half $\tau$-plane by
fundamental domains of $\Gamma_{\theta}$.  b) The subtessellation of the
upper half plane corresponding to the coset $\Gamma_{\theta}/\Z$ (dark lines).
This is the phase diagram of three-dimensional supergravity.  The
tessellation by $\Gamma_{\theta}$ (light lines) is shown to guide
the eye.} {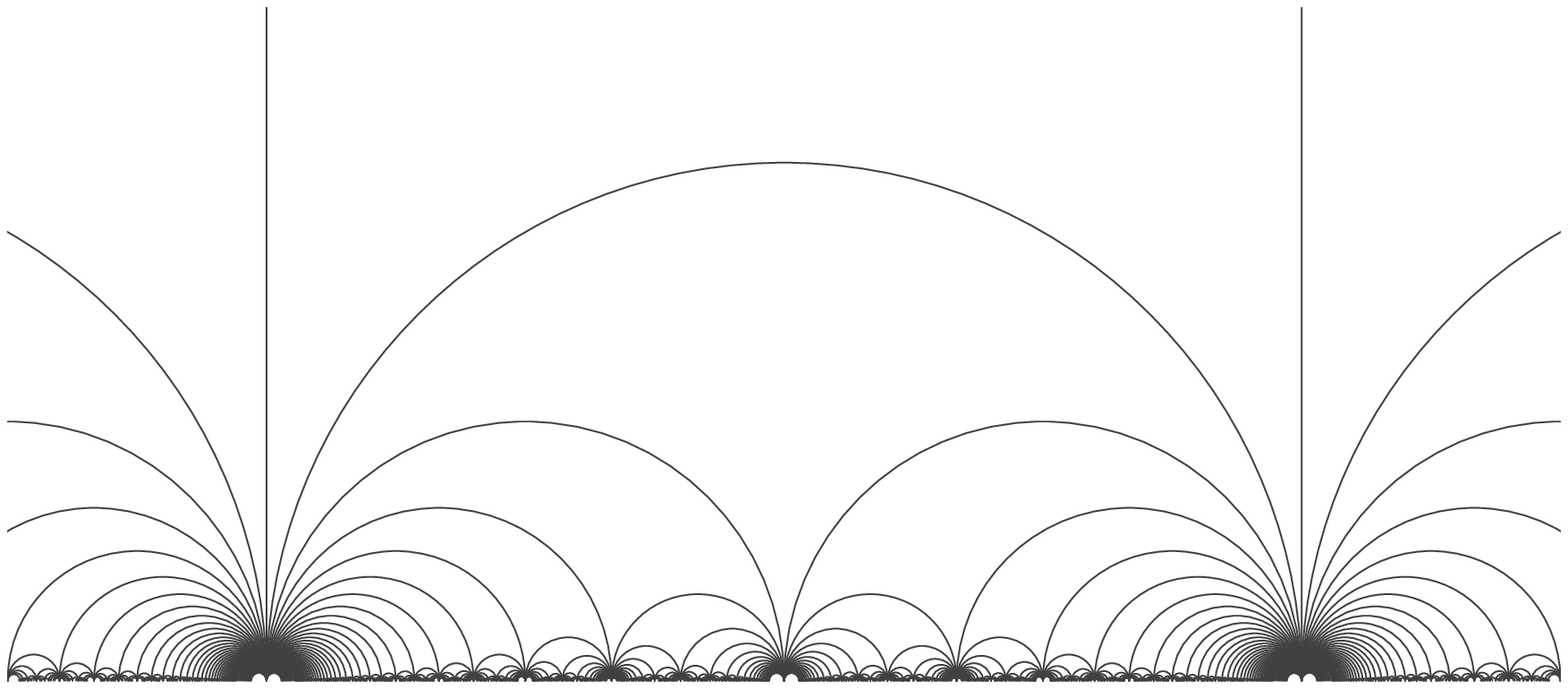}{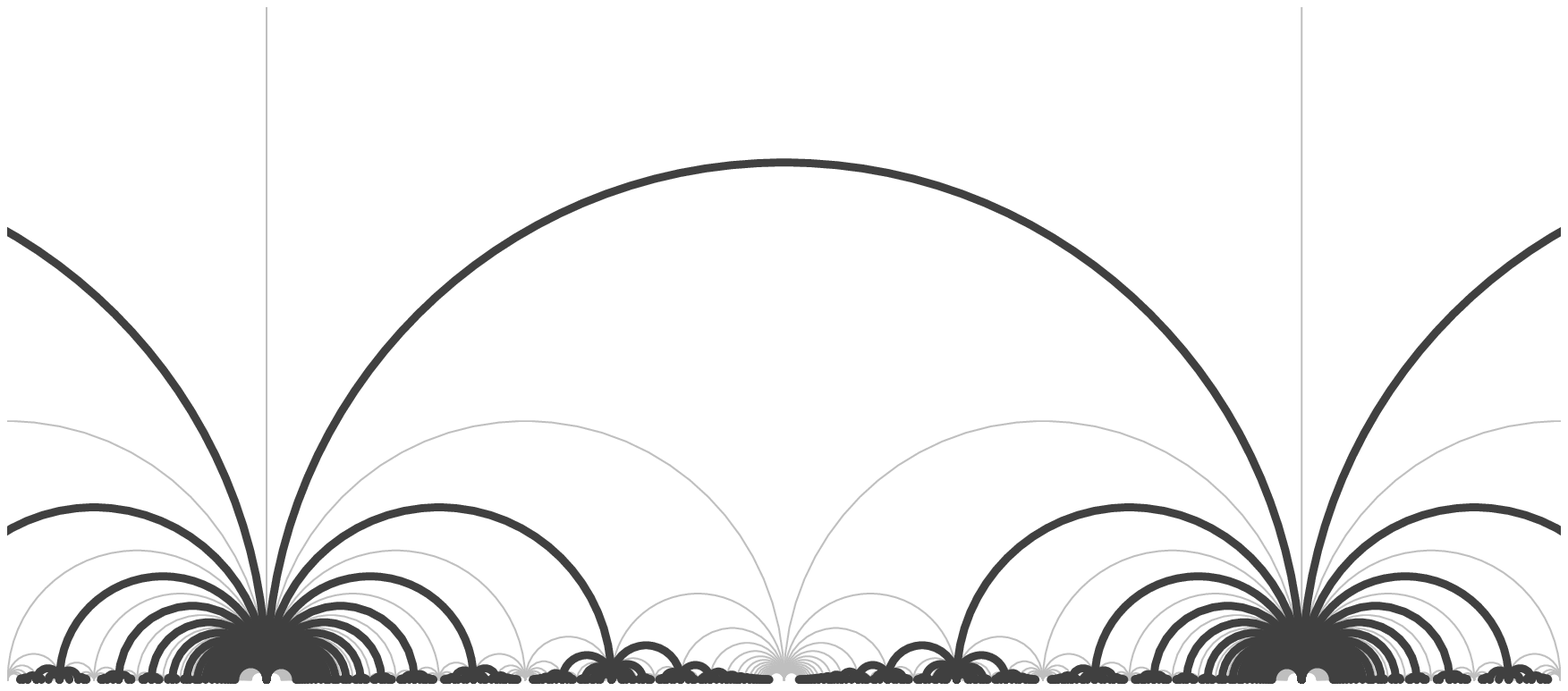}{5cm}{3.3cm}

Within this modular group $\Gamma_{\theta}$, there is a subgroup
isomorphic to $\Z$ generated by the translation $\tau\to\tau+2$.
This $\Z$ is the intersection of $\Gamma_{\theta}$ with the mapping
class group of the handlebody $M_{0,1}$, so modular transformations
in $\Z$ do not generate new handlebody contributions to the
supergravity partition function.  There is a unique saddle point
contribution to $F(\tau)$ for each element of the coset
$\Gamma_{\theta}/\Z$.  The subtessellation of the upper half plane
${\eusm H}$ associated to this coset $\Gamma_{\theta}/\Z$ is shown in Fig. 4b).
As one moves between tiles in this diagram, we expect a first
order Hawking-Page phase transition as different saddles become
dominant.  So the subtessellation depicted in Fig. 4b) should be
interpreted as the phase diagram of three dimensional supergravity.

In \WittenKT, holomorphically factorized partition functions for
three-dimensional gravity were considered.   Numerically, it appears
that\foot{Here we assume the numbers of Ramond and Neveu-Schwarz
primary states at the black hole threshold to be small. In
\WittenKT, there was no natural way to determine these numbers.} the
zeroes of these partition functions lie on the dark curves in Figure
4b), and become dense on these curves in the large $k$ limit
\kanekonote. This indicates that the Hawking-Page transition in
supergravity occurs by the mechanism of Lee-Yang and Fischer, as in
the bosonic case described in Sec. 6. It may be possible to prove
this by extending the analytic arguments of \AKN\ and Sec. 6.

\centerline{\bf Acknowledgments} We are very grateful to R. Canary, D. Gaiotto, M.
Kaneko, J. Maldacena, M. Mirzakhani,  R. Rabadan, N. Seiberg, X. Yin, and
especially P. Sarnak for useful conversations.

\listrefs

\end